\documentclass{aastex61}
\pdfoutput=1 %for arXiv submission
\usepackage{amsmath,amstext,mwe,esint}
\usepackage{soul}
\usepackage[T1]{fontenc}
\usepackage{apjfonts} 
\usepackage[figure,figure*]{hypcap}
\newcommand{\dem}{$DEM(T)$}
\newcommand{\alf}{Alfv\'{e}n}
\newcommand{\omgp}{$\omega_p$}
\newcommand{\Tb}{$T_B$}
\newcommand{\Rsun}{$R_\odot$}
\newcommand{\ether}{$\langle E\rangle$}
\newcommand{\MYhref}[3][blue]{\href{#2}{\color{#1}{#3}}}

\shorttitle{Multi-waveband study of a loop heating event}
\shortauthors{Mohan, A. et al.}

\begin{document}

\title{A weak coronal heating event associated with periodic particle acceleration episodes}
\correspondingauthor{Atul Mohan}
\email{atul@ncra.tifr.res.in}

\author[0000-0002-1571-7931]{Atul Mohan}
\affiliation{National Centre for Radio Astrophysics - Tata Institute of Fundamental Research,Pune 411007, India.}
\author{Patrick I. McCauley}
\affiliation{School of Physics, University of Sydney, Sydney, NSW 2006, Australia}
\author{Divya Oberoi}
\affiliation{National Centre for Radio Astrophysics - Tata Institute of Fundamental Research,Pune 411007, India.}
\author{Alpha Mastrano}
\affiliation{School of Physics, University of Sydney, Sydney, NSW 2006, Australia}

\begin{abstract}
Weak heating events are frequent and ubiquitous in solar corona. They derive their energy from the local magnetic field and form a major source of local heating, signatures of which are seen in EUV and X-ray bands. Their radio emission arise from various plasma instabilities that lead to coherent radiation, making even a weak flare appear very bright. The radio observations hence probe non-equilibrium dynamics providing complementary information on plasma evolution.  However, a robust study of radio emission from a weak event among many simultaneous events, requires high dynamic range imaging at sub-second and sub-MHz resolutions owing to their high spectro-temporal variability.  Such observations were not possible until recently. {This is among the first spatially resolved studies of an active region loop hosting a transient brightening (ARTB) and dynamically linked to a metrewave type-I noise storm. 
It uses imaging observations at metrewave, EUV and X-ray bands, along with magnetogram data. 
We believe this is the first spectroscopic radio imaging study of a type-I source, the data for which was obtained using the Murchison Widefield Array.} We report the discovery of 30 s quasi-periodic oscillations (QPOs) in the radio light curve, riding on a coherent baseline flux. The strength of the QPOs and the baseline flux enhanced during a mircoflare associated with the ARTB. Our observations suggest a scenario where magnetic stress builds up over an \alf\ timescale (30s) across the typical magnetic field braiding scale and then dissipates via a cascade of weak reconnection events.
 \end{abstract}


\section{Introduction}\label{intro}
Solar corona is teeming with weak particle acceleration events. 
These events derive their energy from the local magnetic fields and eventually dissipate as heat.
There is a large body of observational evidence to substantiate that the number of weaker events is much larger than those with higher energy \citep[e.g.][]{Drake71_SXR_characters,Lu91_flaredistribution,crosby93_Xrayflarestats}.
This can easily be seen in the full Sun integrated X-ray light curves. 
Of the 338,661 X-ray flares recorded by the Geostationary Operational Environmental Satellite (GOES) over a period of 37 years, \citet{ash2012_flarestats} showed that about $90$\%  of the flares belonged to classes with energy less than a C class flare ($< 10^{-6}\ W\ m^{-2}$), and only $0.07$\% flares belonged to the strongest GOES category ($> 10^{-4}\ W\ m^{-2}$). 
While not responsible for the steady heating of the quiet corona, these weak flaring events form a significant source of heat flux in their vicinity.
From the brightest to the faintest, the flare energies are known to vary by at least nine orders of magnitude.
Our ability to detect weaker flares is constrained by instrumental limitations. 
Studying these weak flares is essential for understanding the processes which convert magnetic energy to thermal in the solar atmosphere. 
There is also the interesting possibility that these flares are higher energy analogues of those which give rise to the steady coronal heating. 

Metrewaves provide a very interesting and complementary observational window for studying such weak flares.
The electron beams accelerated by these weak flares eventually radiate via resonant wave-particle and wave-wave interactions, leading to coherent plasma emission at \omgp\ or its harmonic \citep{ginzburg1958,Tsytovich69,melrose1972,Melrose09_CoherentEmiss}.
Hence, even energetically weaker flares tend to give rise to very high brightness temperatures making it easier to probe a much weaker population of flares than is currently feasible at EUV and X-rays \citep{ crosby97_radioburst_noXraykin, ramesh13_picoflare_discovery, alissa2015}.
The \omgp\ in the low coronal regions lies in the metrewave band \citep{newkirk1961, zucca2014}.
As the plasma emission is believed to be at \omgp\ (or its harmonic) the frequency of emission can be mapped directly to the local plasma density. 
In the corona, \omgp\ is a function of height, so the observation frequency can be used to infer an approximate height of the emission and probe coronal heights inaccessible to observations at high energies.
These observations of nonthermal emissions probe independent aspects of evolution of the flare plasma in contrast to the observations at high energy bands, which are sensitive to signatures of thermal emission.
Together, they help build a more complete picture.
Another aspect of these coherent emissions is their rapid temporal and spectral variability, as evidenced by numerous prior observations \citep[e.g.][]{wild1957_burstClassification,ellis69_typeIIIb, delanoe1975_typIIIbcharacterisation,mugundhan2017_dnn_est,Suresh17_waveletbasedstudy,sharykin2018_LOFAR_dnn_withtypIIIb}.
However, for a source like the Sun, which has a wide variety of structures spanning a large range in angular size and intrinsic brightness, high fidelity imaging with sufficient time and frequency resolution has traditionally remained challenging.

Snapshot spectroscopic imaging with sub-second, sub-MHz resolutions at metre-wavelengths has now become possible with the advent of new generation low frequency radio interferometric arrays. 
These include the Murchison Widefield Array (MWA; \citealp{Lonsdale09, Tingay2013}), the LOw Frequency ARray (LOFAR; \citealp{vanHaarlem13}) and the Long Wavelength Array (LWA; \citealp{Kassim2010}).
The MWA can provide images with resolutions down to $0.5\ s$ and $40\ kHz$.
The design of MWA is especially well suited for high Dynamic Range (DR) imaging.
Using an unsupervised imaging pipeline, it has recently been demonstrated that the DRs of MWA solar images typically exceeds $10^3$  \citep{Mondal19_AIRCARS}. 
The lowest DRs of a few 100 are obtained under the most challenging circumstances of a quiet featureless Sun; and under favourable conditions DRs exceeding $10^5$ have been obtained.
This study is an example of the kind of novel explorations enabled by the availability of these high DR images. 
Till recently, most solar radio studies were largely limited to studying the dominant emission feature on the Sun.
The high dynamic ranges of these radio images now allow us to effectively isolate the emissions from  different features simultaneously present on the Sun, of potentially very different strengths, and carry out detailed studies of weak emission features.
\citet{Atul19_QPO} provides a recent example of such studies which combines snapshot spectroscopic imaging with high energy observations and magnetic field modelling to investigate the dynamics of a weak active region jet associated with groups of type-III radio bursts.

{Here we present the first spectroscopic radio imaging study of a type-I noise storm associated with a weak active region transient brightening (ARTB).}
Simultaneous observations from the metrewave, EUV and X-ray bands are used alongside concurrent vector magnetogram data.
A salient feature of this work is that it used multi-waveband light curves constructed using images of the regions of interest. 
{Earlier studies of ARTBs have mainly been restricted to high energy bands which demonstrated the link between these events and microflares \citep[e.g.][]{shimizu92ARTB_yohkoh,Berghmans99_ARTB_EIT,warren07_ARTB_hinode}. 
Using imaging observations from the 1 -- 18 GHz band, a link between microwave bursts and ARTBs has also been demonstrated  \citep{gopalswami94_ARTB_microwavesource,white95_ARTB_nobeyama,Nindos99_ARTB_microwave}. 
In metrewave bands, 
some studies have shown that the transient EUV brightening events could produce accelerated particle beams that caused impulsive metric type-III radio bursts \citep{kundu86_ARTB_mwaveTypeIII,alissa2015}.
{For metric type-I noise storms the higher energy (EUV and X-ray) have remained elusive and their physical drivers are also not well understood  \citep{elgaroy77_noiseStormBook, white2007_radiobursts}. However, they had been long associated with active region magnetic field structures and sunspot groups \citep{wild56_noisestorm_prominance_Connection,wild1970}. Most of the noise storms were associated with flaring events, and were interpreted as nonthermal emission triggered by flare-accelerated electrons \citep[e.g.][]{dodson58_typeI_flare_statsLink, lesqueren64_flare_typeINRH169locs, willson98_CME_74MHztypeI,kathiravan07_stats_CMEdrivenTypeI,Iwai11_CME_enhancetypeI_link}. Some of these bursts also show quasi-periodic brightness fluctuations ranging from a few seconds to minutes in their light curves \citep[e.g.][]{wild51_slowlyVar_QPOs,sastry69_secQPOs_typeI, Sundaram04_QPO_110min_noisestormbgd}, the origins of which and possible links to high energy sources are not well understood. Occasionally, these sources are observed associated with non-flaring active regions \citep{smith62_flare_typeI_correlStats}, implying that flaring is not a necessary condition for the existence of these events. In a case study \cite{iwai12_typeI_smallscaleEUVmagDyn_link} demonstrated a temporal correlation between the onset of such type-I bursts in the full-sun integrated dynamic spectrum and small scale EUV and magnetic field activities at a non-flaring active region. However, the dynamical/causal link between the type-I burst and small scale EUV and magnetic activities still remained unclear until \citet{Lin17_typeI_smallEUVflare_link} demonstrated a good correlation between spatially resolved light curves of type-I burst sources at three frequencies and the co-spatial EUV and magnetic activities.}
This work presents a detailed investigation of an ARTB -- type-I system, with the aim of understanding the nature of emission at high energy and metrewave bands, the dynamical correlations which exist between them, and their implications for the physical mechanism involved. 
}
%This work simultaneously addresses the aforementioned issues pertaining to both ARTBs and type-I sources.}
%in a particular case, using a novel multi-waveband observational approach assisted by magnetic field modelling.}

The radio data used in this study come from the MWA; the EUV data from the Atmospheric Imaging Assembly (AIA; \citealp{Lemen2012}) on board Solar Dynamics Observatory (SDO); the X-ray data from the Reuven Ramaty High Energy Solar Spectroscopic Imager (RHESSI; \citealp{lin2002}); and the
vector magnetogram data used for magnetic field modelling came from the Helioseismic and Magnetic Imager (HMI; \citealp{Scherrer12}) onboard the SDO.
The analysis of these spatially resolved light curves led to interesting insights about the evolution of physical parameters of the system during the ARTB driven microflare, as it progress from its pre-flare state to the heating phase followed by the cooling phase.
In particular, the ability to detect fine emission features in SPatially REsolved Dynamic Spectra\footnote{Analogus to a dynamic spectrum, SPREDS is the dynamic spectrum of the emission arising from a specific region on the Sun.} (SPREDS; \citet{Atul17}) allowed us to detect quasi-periodic oscillations in the radio emission associated with the type-I source. We also suggest a physical scenario to explain these observations.

The organisation of the paper is as follows: the event and the multi-waveband observations are described in Section \ref{obs}. The radio, X-ray, EUV and HMI data analyses are described in Section \ref{analysis}. Section \ref{discussion} presents a discussion of the observations, the inferences which can be drawn from them, with their implications and likely interpretations.
Section \ref{conclusion} presents the conclusions.
\begin{figure} 
  \centering
  \includegraphics[width=17cm,height=12cm]{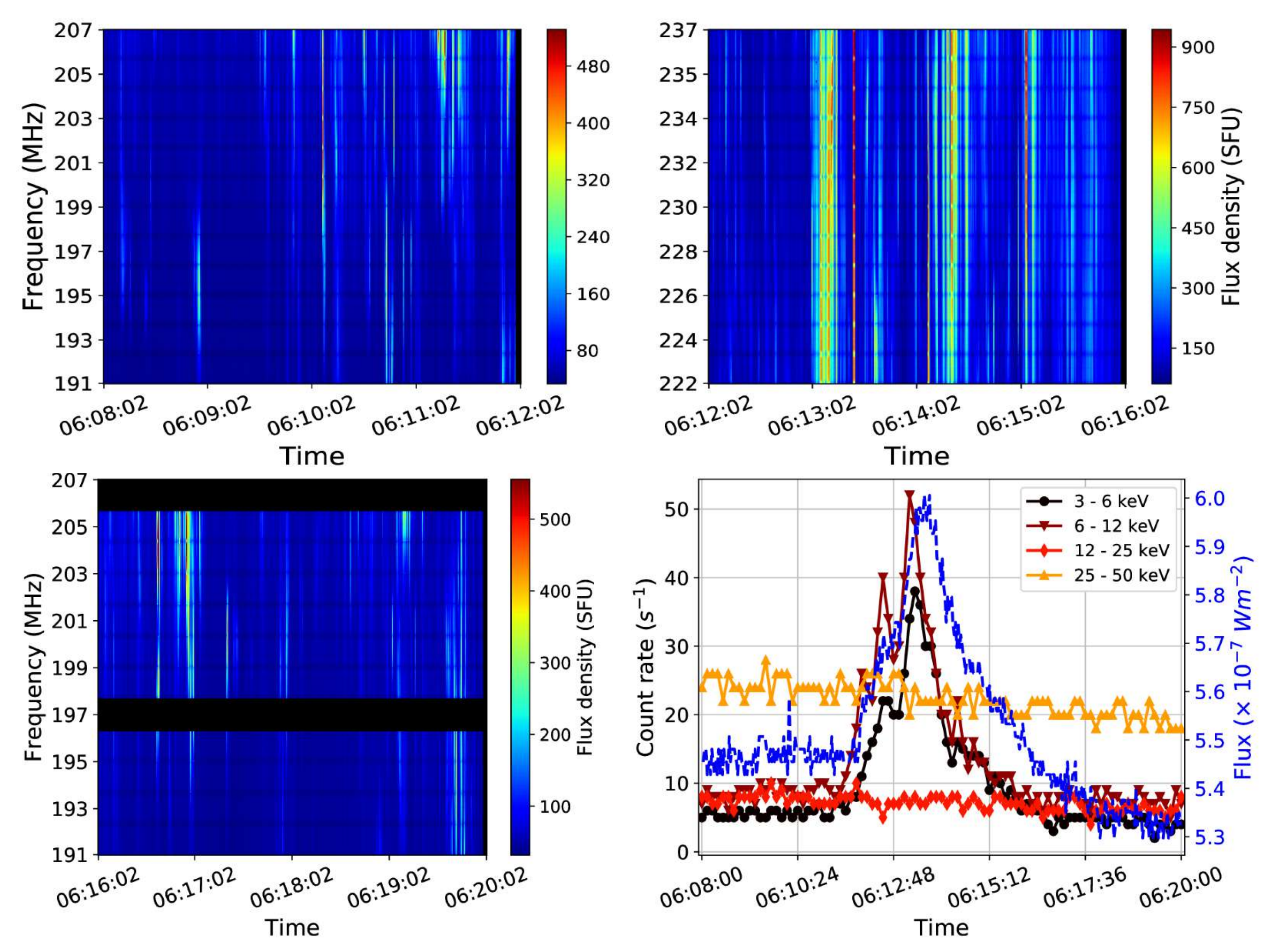}
 \caption{{\it Top:} {The full sun integrated }dynamic spectrum from the MWA during the first 8 minutes of observation. {\it Bottom left: }{Full sun integrated MWA dynamic spectrum for the last 4 minutes of observation.} Black bands denote bad data. {\it Bottom right: } RHESSI X-ray flux in the various energy bands. GOES 1 -- 8 \AA\ light curve is shown in dashed blue line.}
  \label{fig1}
\end{figure}
\begin{figure}
  \centering
  \includegraphics[width=14cm,height=13cm]{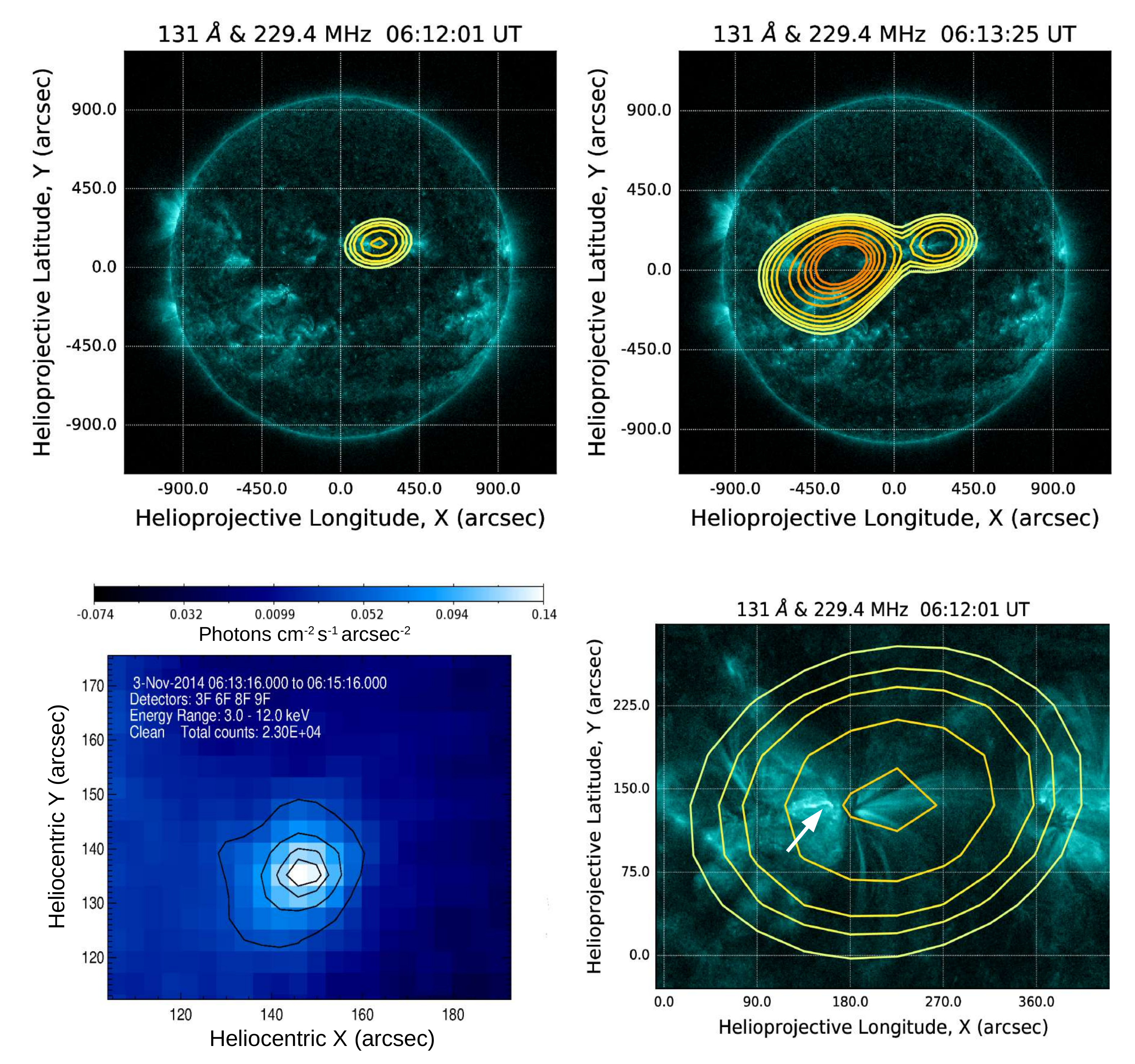}
  \caption{{\it Top: }AIA 131 \AA\ images before ({\it left}) and during ({\it right}) the type-III burst event. Overlaid are simultaneous MWA 229.4 MHz contours. Contours are marked at absolute \Tb\ levels of .5, .7, .9, 1.3, 1.8, 3.6, 5.4, 7.2, 9 and 11 $\times 10^8 K$ for all radio images in the figure. The radio source common in the two panels is a persistent source. A stronger source appeared in the eastern side of the disk during the burst. {\it Bottom: } Left panel shows a RHESSI image of the flare source with contours marked at 30, 60, 80 \& 90 \% of the peak. Right panel zooms into the image in the top left panel, revealing the association of the persistent radio source to a bright active region loop. The arrow points at the centroid of the RHESSI source shown in the left panel.}
 \label{fig2}
\end{figure}
\section{Observations}\label{obs}
The radio data were gathered using the MWA on November 3, 2014 between 06:08:02 -- 06:20:02 UT under the observing code G0002. The observations were made across a bandwidth of 15.36 MHz with a time and frequency resolution of 0.5 s and 40 kHz, respectively. The data comprise three contiguous observations of four-minutes duration with neighbouring central frequencies. The first (06:08:02 -- 06:12:02 UT) and the last observations (06:16:02 -- 06:20:02 UT) were centred at 199 MHz. The intervening observation (06:12:02 -- 06:16:02 UT) was centred at 229 MHz.

The day was characterised as a period of medium activity, with 10 type-III radio bursts, 8 type-IV radio bursts and a GOES class C1.4 flare\footnote{\MYhref{ http://www.solarmonitor.org}{ http://www.solarmonitor.org}}. The Space Weather Prediction Center (SWPC) reported groups of type-III bursts from 06:02 -- 07:22 UT which includes the period of MWA observations\footnote{\MYhref{ftp://ftp.swpc.noaa.gov/pub/indices/events/README}{ftp://ftp.swpc.noaa.gov/pub/indices/events/README}}. The 1 -- 8 \AA\ GOES band shows evidence for a B6 class flare. The X-ray light curves from RHESSI show signatures of a weak flare in the 3 -- 12 keV bands. This event was not reported in the SWPC event list, likely because its flux did not sufficiently exceed the background. Figure \ref{fig1} shows the MWA dynamic spectra for the three observations. The bottom right panel of the figure shows the GOES and the RHESSI X-ray light curves for the different energy bands during the observation period.

The top panel of Figure \ref{fig2} shows AIA 131 \AA\ images before (left) and during (right) the type-III bursts. 
Overlaid on each of them are the MWA 229 MHz contours at the respective times. The bright radio source in the north-west part of the visible solar disk is a persistent source present throughout the observation.
The image in the top right panel shows a brighter radio source located in the eastern half of the disk. This source consistently appears during every type-III burst with high brightness temperatures (\Tb\ $\approx 10^8$ -- $10^9$ K). The darker \Tb\ contours signify the enhanced brightness of this source relative to the persistent radio source.
Six groups of type-III bursts took place during the observation period and were spatially and temporally associated with a weak active region jet seen in all observing bands of the AIA. 
A detailed study of the type-III bursts and the associated jet is presented in \cite{Atul19_QPO}. Here we focus on the persistent radio source. 
The bottom-left panel of the figure shows a RHESSI image made by averaging the data over 3 -- 12 keV and 120 s (06:13:16 -- 06:15:16 UT).
We find a bright compact X-ray source centred at ($150^{\prime\prime},135^{\prime\prime}$), close to the location of the persistent radio source. An X-ray source of similar {(flux) strength} was found to be associated with the type-III burst sources as well. So the X-ray flare seen in GOES and RHESSI data (Fig. \ref{fig1}) has contributions from both these sources. The bottom-right panel zooms into the north-west region of AIA 131 \AA\ map corresponding to the location of the persistent radio source. Overlaid are the MWA 229 MHz \Tb\ contours. The location of the associated RHESSI source is marked by an arrow. An EUV bright active region loop is associated with the radio and X-ray sources. This loop was persistently bright during the observation period. 
{It belonged to an active region numbered AR 12203 by National Oceanic and Atmospheric Administration (NOAA). No flares had been recorded by NOAA from this active region on this day. But, this region had the largest number of sunspots (9) and sunspot area (150 millionths) among all the 10 active regions recorded on that day.}

We note that the radio source is offset from the location of the bright EUV loop.
This is expected as the radio source is located much higher up in the corona and its location is determined by magnetic and density fields in the region. The meterwave radiation from the burst source is also subject to propagation effects like scattering, refraction and ducting due to coronal plasma structures and turbulence in the medium (\citealp[e.g.][]{steinberg1971,Arzner1999, kontar2017, Atul19_QPO}). Simulations have shown that these effects can shift the location of the radio source by a few arc minutes (\citealp{robinson_scat1983, Wu2002}). 
The bright active region loop is the only EUV and X-ray source seen in the vicinity of the radio source. 
Hence, despite the offset, it is reasonable to associate the radio source with it.
This system is the focus of the present study. 
%\citet{white2007_radiobursts} reports that groups of type-III bursts usually take place 
%during the rising phase of the GOES X-ray flare and cease close to the X-ray peak.
%Also, \citet{Reid17_typIII_Xray_Corr} find that only about 28\% of the type-III bursts in their ten year study had a corresponding RHESSI X-ray flare associated with them. Together, these facts imply that the type-III radio source and X-ray source are not related and their temporal overlap is a chance coincidence.

\section{Analysis}\label{analysis}
As mentioned in Section \ref{obs}, the meterwave radio emission is generated at a much greater height in the corona as compared to the bright active region loop. To get an estimate for the radio source height we use the fact that the coronal plasma frequencies fall in metre-wavelengths. Being proportional to the square root of local density that falls monotonically with heliocentric distance, the local plasma frequency in the corona decreases with height in the corona. Since any radiation produced at a frequency less than the local plasma frequency gets efficiently absorbed, the observed frequency of radiation must be generated at a height greater than the region where local plasma frequency equals the observation frequency. This provides a straight forward way to estimate a lower limit to the height of radio emission using a coronal electron density model.
Using the model provided by \citet{zucca2014} (hereafter the Z model), we estimate the source height limit to be $1.14$ \Rsun. 
Assuming a semi-circular geometry for the bright AIA loop and using observed separation of footpoints as its diameter, we estimate a height of $\approx 1.02$ \Rsun  for this loop.
The radio and the EUV emission thus comes from regions well separated in height.
 
This section is divided into four subsections, describing the analyses of EUV images from the AIA, radio observations from the MWA, the RHESSI imaging data, and vector magnetogram from the HMI, respectively.

\subsection{AIA data analysis}
\begin{figure}
  \centering
  \includegraphics[width=15.6cm,height=6cm]{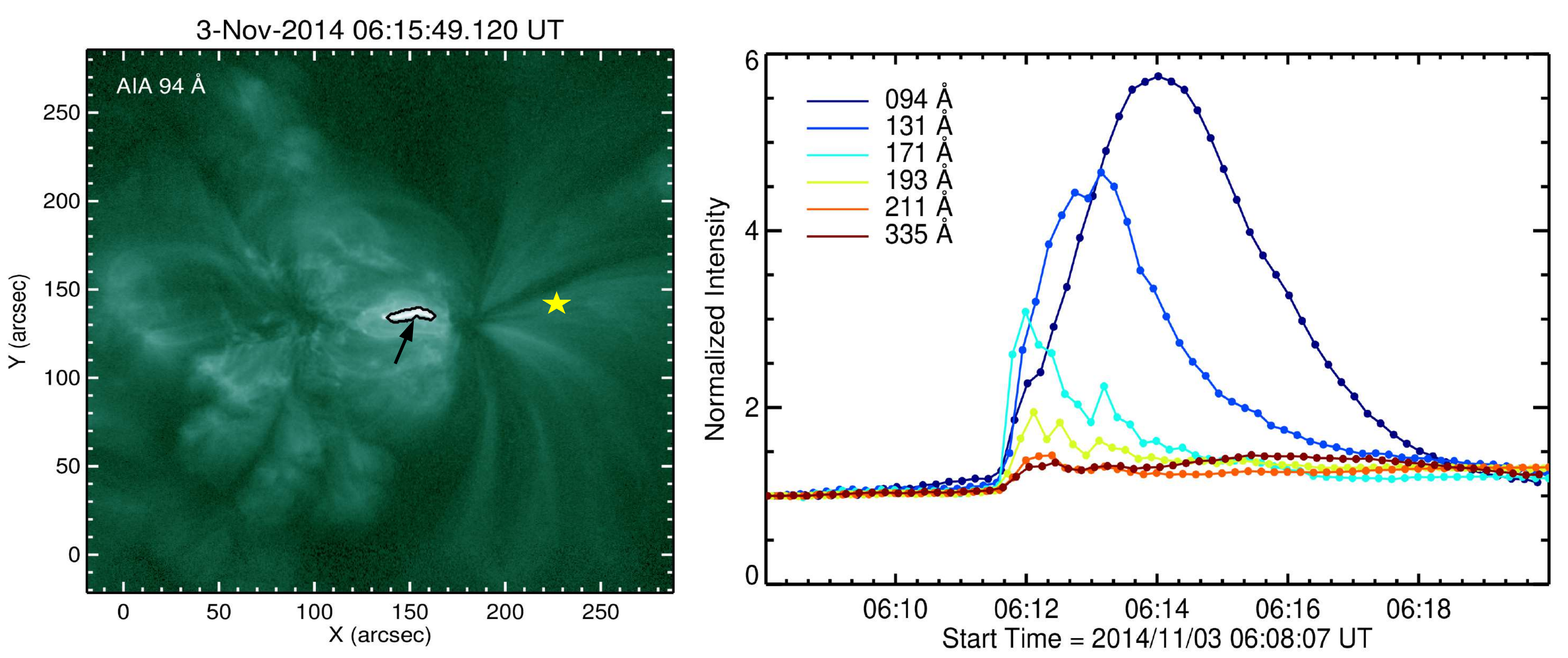}
  \caption{{\it Left: }The active region hosting the bright loop. A boundary is defined for the bright loop region, marked in black, based on a criteria defined in the text. Yellow star on the right identifies the large loop structure on which the radio contours overlay. The black arrow points to the peak flux location of the RHESSI source. {\it Right: } Light curves of the bright loop region in various AIA bands sensitive to coronal temperatures, within the defined boundary}
  \label{fig3}
\end{figure}

The bright active region loop seen in the coronal EUV bands of the AIA, namely 94, 131, 171,193, 211 and 335 \AA\ were analysed using routines in the SolarSoft Ware (SSW; \citealp{freeland1998_SSW}) package. 
Level 1 data were obtained from the Virtual Solar Observatory (VSO) and co-aligned using the SSW routine \texttt{aia\_prep}. 
The left panel of Figure \ref{fig3} shows a 94 \AA\ image of the active region containing the bright loop. 
The black boundary around the loop defines the region used for further analysis. This boundary was defined by thresholding the image above 40 times the mean value and slightly dilating the resulting mask to encompass fully the region that underwent significant brightening during the event. 
The light curves in the right panel of Figure \ref{fig3} span the duration of MWA observations and correspond to the average value within this boundary, normalized by the intensity at the beginning of the time series, before the onset of the event. 
A clear rise in the normalised intensity is observed starting around 06:11:40 UT. The intensity in different bands peaks at different times with the 94 \AA\ light curve peaking after 06:14:00 UT. 
It is difficult to conclude anything about the evolution of physical fields at the loop region by just examining these light curves. This is because of two reasons - the observed intensity is a function of temperature and density; and the AIA bands have very broad and overlapping temperature response functions. To study the evolution of temperature and density, we use a technique based on differential emission measure \citep{withbroe75_DEM_technique} by combining light curves from all the AIA bands. 
\subsubsection{Differential Emission Measure ($DEM(T)$)}\label{DEM}
The observed intensity in the various AIA bands is the sum of emission from multi-thermal plasma components along the line of sight. 
The Differential Emission Measure ($DEM(T)$) is a quantity that characterises the individual contributions from these different temperature ($T$) components. 
By estimating $DEM(T)$ at various instances of time during our observation, we can detect signatures of heating by looking for a rise in the high temperature component. Using the \dem\ function we can obtain the total Emission Measure ($EM$) by simply integrating over \dem. Assuming that the number density of electrons ($n_e$) and protons ($n_H$) are the same, the observed intensity at a wavelength ($\lambda$), $I_\lambda$, $EM$ and \dem\ are related as,
\begin{eqnarray}
    I_\lambda &=& \int [{n_e}^2 \frac{dz}{dT}](T) \Lambda(T,n_e,\lambda,\chi) R_{\lambda} dT \label{eqn1},\\
    DEM(T) &=& {n_e}^2\frac{dz}{dT} \label{eqn2},\\
    EM &=& \int DEM(T) dT = \int {n_e}^2 dz\label{eqn3},
\end{eqnarray}
where dz is a height segment present along the line of sight at a particular temperature, $R_{\lambda}$ is the known instrumental response function, and $\Lambda$ is the contribution function that gives the emissivity of the plasma at a given $T$, $n_e$, and wavelength assuming a typical ion abundance of $\chi$ in the corona. 
$\Lambda$ can be computed for any $n_e$ and $T$ using the CHIANTI database of atomic spectral lines (\citealp{dere97_chianti,landi99_chianti_releaseIII}). 
$I_{\lambda}$ is the observable in different AIA bands. Thus each AIA bands contribute one equation corresponding to Equation \ref{eqn1} giving rise to a system of equations which can be solved to obtain \dem. 
\begin{figure}
  \centering
  \includegraphics[width=17cm,height=5.5cm]{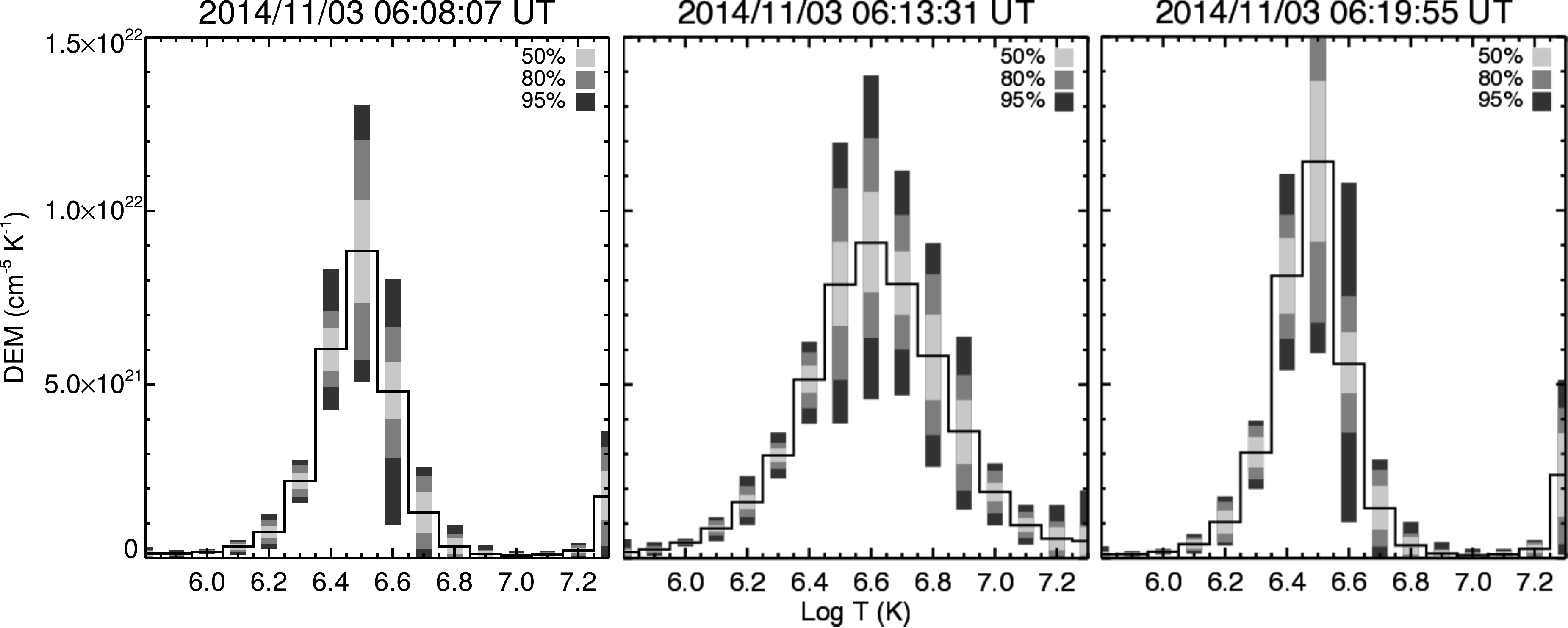}
  \caption{\dem\ for the loop region during three stages of the brightness enhancement event: before, around the maximum and after (left to right). 100 Monte Carlo runs were carried out to estimate the \dem, by randomly varying the input data each time within the uncertainty region.The shaded grey bars in the plot represent the regions that encompass 50\%, 80\%, and 95\% of the solutions for a given temperature bin.}
  \label{fig4}
\end{figure}

We integrate the \dem\ to obtain the total EM and DEM-weighted average temperature ($\langle T \rangle$) along the line of sight (LOS). By assuming a LOS depth $D$, we can then estimate the mean electron density ($\langle n_e \rangle$) and mean thermal energy ($\langle E \rangle$) for the emission region:
\begin{eqnarray}
    \langle T\rangle &=& \frac{\int DEM(T) T dT}{EM} \label{eqn4},\\
    \langle n_e\rangle &=& \sqrt{\frac{EM}{D}}\label{eqn5},\\
    \langle E\rangle &=& 3 \langle n_e\rangle k_B \langle T\rangle V \label{eqn6},
\end{eqnarray}
where $k_B$ is the Boltzmann constant and $V$ the volume of the loop. For the chosen loop region, we estimated $DEM(T)$ at each 12-sec time step between 06:08:00 -- 06:20:00 UT. 

\begin{figure}
  \centering
  \includegraphics[width=9cm,height=10cm]{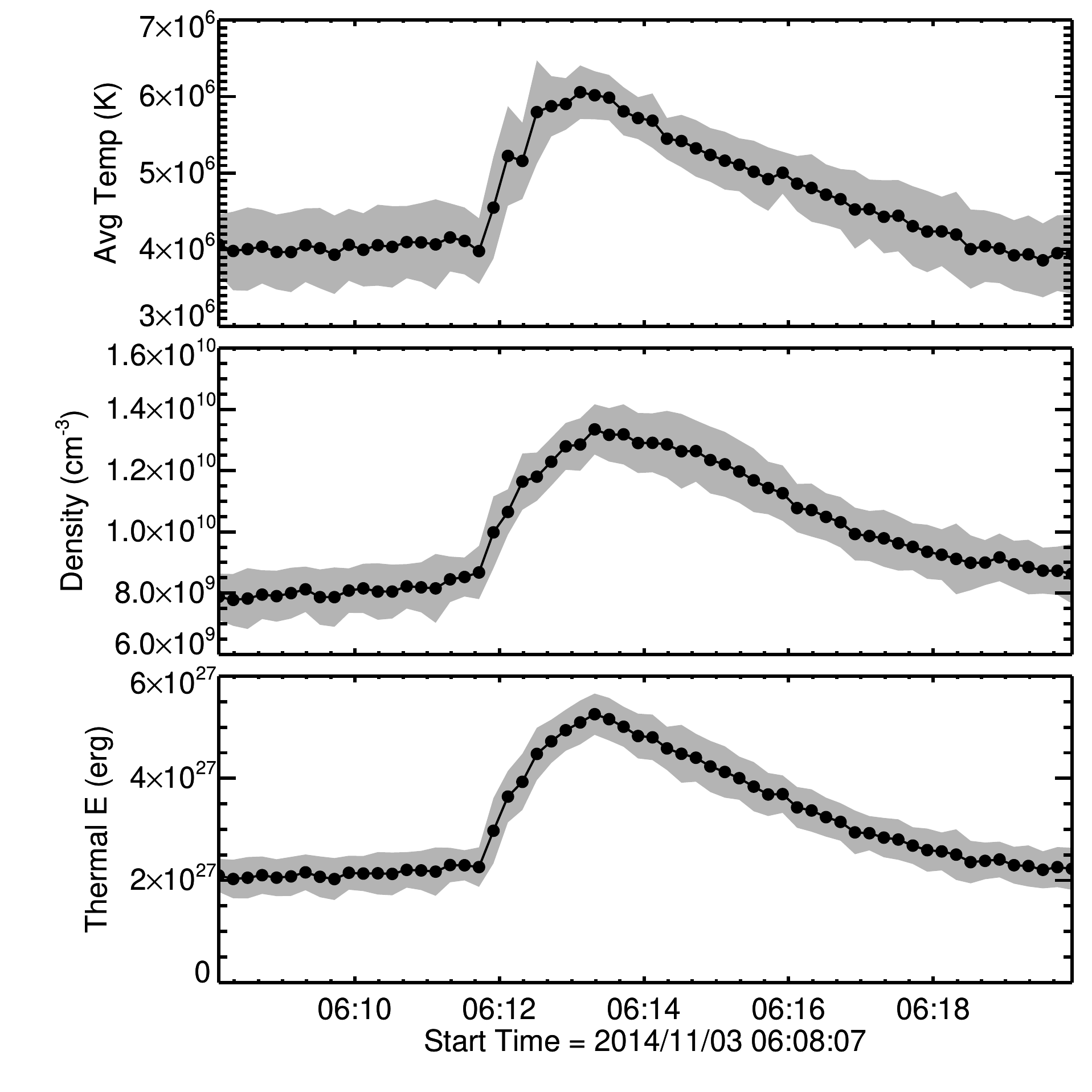}
  \caption{The evolution of \dem\ weighted temperature, mean density and mean energy of the event. The shaded regions denote the standard deviation of each parameter across the 100 Monte Carlo iterations.}
  \label{fig5}
\end{figure}
To estimate \dem, we used the \texttt{xrt\_dem\_iterative2} algorithm available in SSW \citep{Weber04_DEM_xrtDemIterative, Golub2004_DEM_method}.
The code was designed for the X-ray Telescope (XRT) onboard 
the Hinode satellite and was later adapted for AIA data (e.g. \citealp{Schmelz11,Cheng12}). 
The approach is to forward-fit the \dem\ by predicting the observed count rates from an initial guess and iteratively modifying the solution to minimize the  $\chi{}^2$ between the observed and the predicted fluxes.
Iterative Levenberg--Marquardt least-squares minimization is used, \dem{} is interpolated between the available points using spline knots, and this method has been shown to reproduce model DEMs of active regions using simulated AIA data \citep{Cheng12}. 
Uncertainties are obtained from 100 Monte Carlo runs, for which the data are randomly varied within their uncertainties prior to each pass through the algorithm.  
We obtained date-dependent AIA response functions from the SSW routine \texttt{aia\_get\_response} using the ``chiantifix'', ``evenorm'', and ``noblend'' keywords. 
Our implementation is adopted from \citet{Reeves15}, but we compute \dem{} over a narrower temperature range of log($T$) = 5.8--7.3. 
This range spans the breadth of the AIA's temperature response peaks, 
and although there is some sensitivity at higher and lower temperatures, the data are not well constrained beyond this range.
Incorporating higher-temperature soft X-ray observations would improve the DEM estimation, but those data are not available for this event. Figure \ref{fig4} shows the \dem\ distribution derived at three epochs; before, during the peak and after the brightening event (left to right). The DEM distribution before and after the loop brightening event look very similar with a peak temperature of $\approx 3$ MK. The \dem\ obtained during the middle of the period when various AIA bands attain their peak intensity (06:13:31 UT), show a significant rise in the high temperature component.

Estimation of $\langle n_e\rangle$ and $\langle E\rangle$ require the depth and volume of the loop respectively. The loop was assumed to be a cylinder with a diameter equal to its observed thickness. This gives it a radius of 7 AIA pixels ($\approx 3$ Mm). The length of the loop segment is about $22$ Mm. The volume of the loop then turns out to be about $1.58 \times 10^{26}$ cm$^{3}$. Note that the loop segment length is a projected value, but it was estimated taking care of the curved structure rather than just the footpoint separation. 
If one assume the loop to be semi-circular and the foot point separation to be its diameter, then the loop length comes to around $32$ Mm. This will not cause large differences to the \ether\ estimates. 
Figure \ref{fig5} shows the evolution of $\langle T\rangle$, $\langle n_e\rangle$ and \ether. The error bars on each quantity is estimated as its standard deviation across the set of solutions obtained from the 100 Monte Carlo runs. The impulsive rise in all of these quantities is evident. This evolution primarily owes to the evolution of the high temperature components in the \dem. Hereafter, this period of evolution of thermal properties of the loop plasma will be referred to as the ``flare". 
{The physical size of the bright loop region, the estimated range of variation of thermal properties and the flare duration during the period of transient EUV brightening, all lie within the typical ranges for an ARTB \citep{shimizu96_artb_review}}. 

\subsection{Radio Analysis}\label{radio_analysis}
The MWA data were imaged at a time cadence of 0.5 s and a spectral resolution of 160 kHz. Imaging was done using an automated radio interferometric imaging pipeline called the Automated Imaging Routine for Compact Arrays for Radio Sun (AIRCARS; \citealp{Mondal19_AIRCARS}), which is optimised for arrays having a dense central core of collecting area, like the MWA. AIRCARS uses imaging and calibration routines from Common Astronomy Software Applications (CASA; \citealp{casa}) and relies on the technique of self-calibration (\citealp{pearson1984, cornwell99}). 

\begin{figure}
  \centering
  \includegraphics[width=18cm,height=6cm]{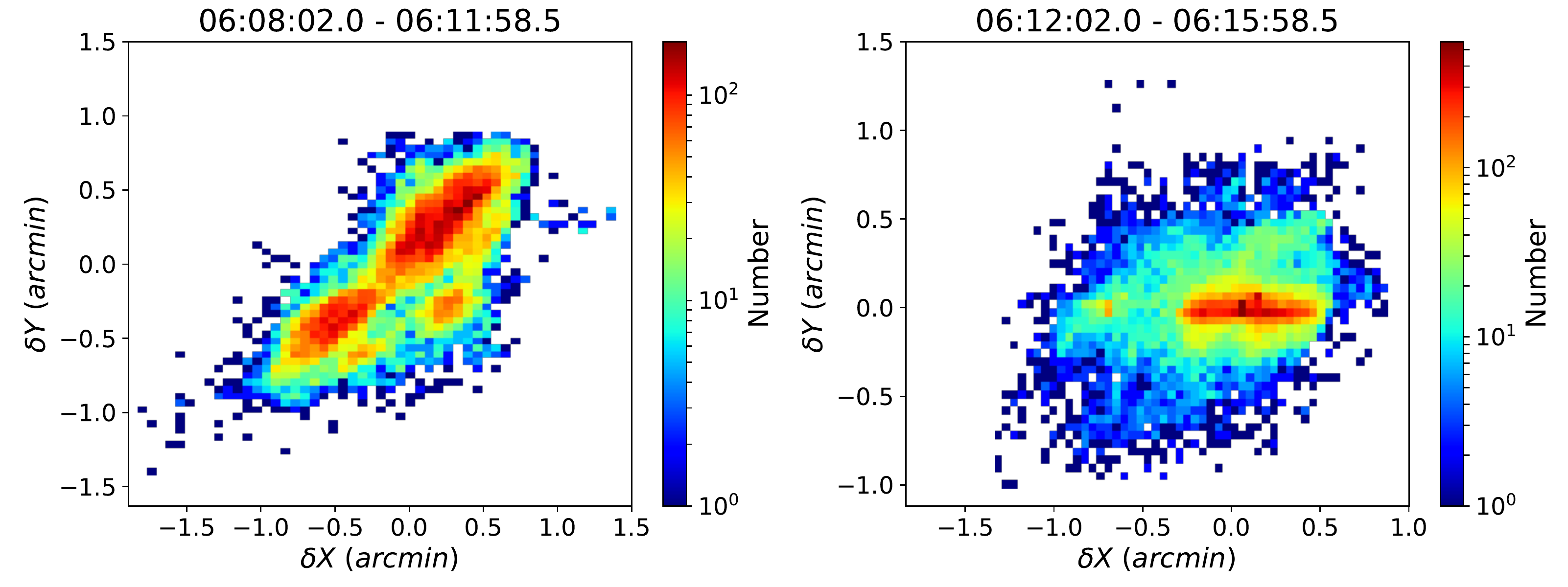}
  \caption{2D histogram of the relative sky location of the radio source with respect to its mean for the first two sets of observations. The observations are centred at 199 MHz ({\it left}) and 229 MHz ({\it right}).}
  \label{fig6a}
\end{figure}

The images used had dynamic ranges (DRs) varying from $\approx 250$ -- $6900$ depending on the solar activity level. The highest DRs were obtained during the times of high integrated flux density. 
During the type-III event, we obtained DRs well exceeding 1000. The \Tb\ of the intermittent type-III source varied between $\approx 10^8$ -- $10^9$ K, while that of the persistent radio emission source ranged from $\approx 5 \times 10^7$ -- $10^9$ K, with a mean \Tb\ well above $10^8$ K. 
The high \Tb\ imply a non-thermal emission mechanism. 
The high DRs in these images allowed us to isolate the emission from the persistent radio source across the entire time and frequency span of our observations, even in the presence of the significantly more intense intermittent burst source. 
{Location of the persistent radio source in the snapshot spectroscopic images was always close to the bright loop, as seen in the example image in Fig. \ref{fig2}. However, its precise sky location fluctuated over the duration and across the spectral band of observation.
To quantify this fluctuation, a 2D Gaussian function was fitted to each image of the radio source across frequency and time, and the sky location of the best fit Gaussian was recorded. 
Good quality fits were obtained for  about 99.5\% of the snapshot spectroscopic images, with a median uncertainty of about 0.01$^\prime$ in the RA and DEC estimates.
Figure \ref{fig6a} shows 2D histograms of the shift in the source sky location from its mean, during the first eight minutes of observations. 
The distribution is clearly asymmetric about the mean location, though the magnitudes of these shifts are typically less than $\pm1^{\prime}$. This is less than 25\% of the typical Full Width at Half Maxima (FWHM) of the synthesised beams at the respective observation frequencies. FWHM of the synthesized beams at 199 MHz and 229 MHz, are (4.4$^\prime$, 3.5$^\prime$) and (4$^\prime$, 3$^\prime$) respectively. }

\subsubsection{SPatially REsolved Dynamic Spectrum (SPREDS)}
\label{subsubsec:SPREDS}
\begin{figure}
  \centering
  \includegraphics[width=14.7cm,height=10.5cm]{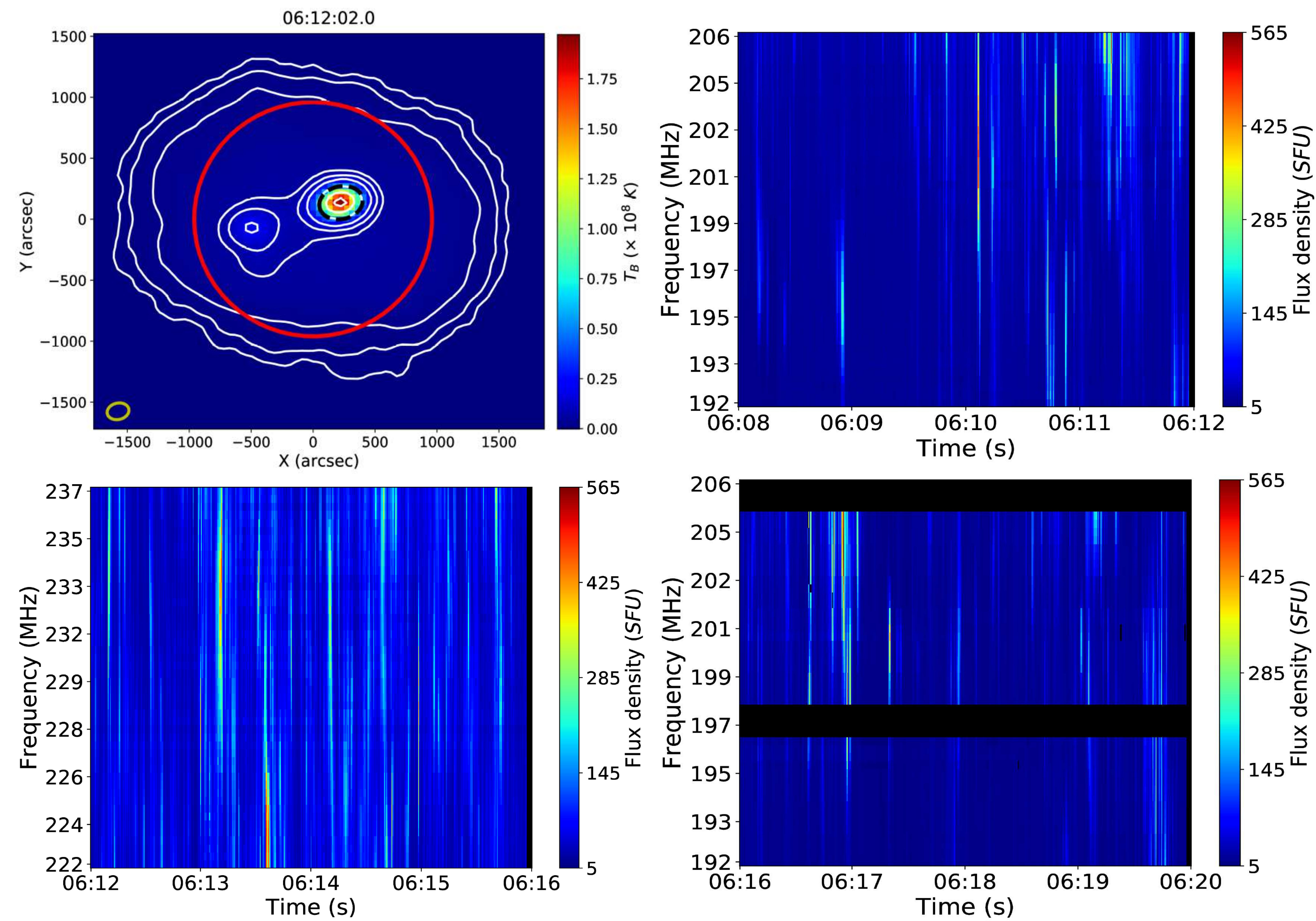}
  \caption{{\it Top left:} A snapshot image of the Sun at 229.36 MHz. Contours mark 0.3, 0.5, 1, 3, 5, 10, 30, 60 and 90 \% of the peak \Tb. 
  The persistent bright radio source chosen for generating SPREDS is marked by a black dashed ellipse. 
  The red circle marks the solar optical disk and the yellow ellipse at the bottom left shows the PSF (FWHM $(4^{\prime},3^{\prime})$). 
  {\it Other panels:} The SPREDS for the persistent source for the three 4-minute observations.}
  \label{fig6b}
\end{figure}

{The SPREDS was obtained by recording the flux density of the radio source across frequency and time, as observed in the snapshot-spectroscopic images.
Top left panel of Figure \ref{fig6b} shows a sample \Tb\ map at 229 MHz and the region used for obtaining the SPREDS is marked in black dashed ellipse. 
This region was centred at the persistent source and defined to have an area twice that of the point spread function (PSF).}
The SPREDS for this region for each of the 4-minute observations are shown in the remaining panels. 
We will use flux density instead of \Tb\ in further discussions as it is physically more meaningful a quantity in the case of non-thermal emission.
Numerous bright emission features of flux density $>\ 200$ SFU (Solar Flux Unit; 1 SFU = $10^{-22}$ Wm$^{-2}$Hz$^{-1}$) are seen in the SPREDS with a typical extent of $\approx 0.5 $ -- $2$ s in time and $\approx 10 - 13$ MHz in frequency during the flare period. The source flux density varies between $\approx 20$ -- $525$ SFU ($\approx 5\times 10^7$ -- $10^9\ K$ in $T_B$) during the observation. 
{The observed emission properties in the SPREDS of the persistent radio source are typical of a type-I noise storm, which comprise numerous fine bursts riding over a nonthermal background emission} \citep[e.g.][]{Elgaroy1970_typIcharacterisation,mclean81_rev_bursttypes}.

\begin{figure}
    \centering
    \includegraphics[width=18cm,height=5.5cm]{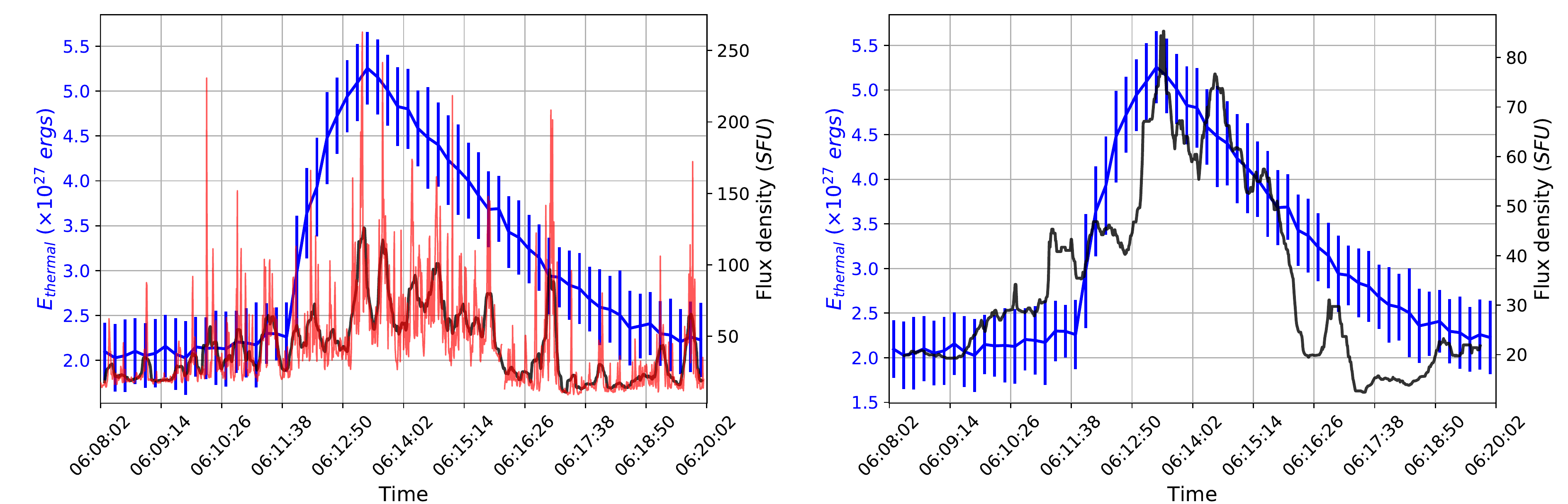}
    \caption{{\it Left: } 
    The thermal energy variation obtained from DEM analysis is shown in blue. 
    The light curve of the radio flux density averaged across the observation band is shown in red. The black line is obtained by applying a 10 s wide running mean filter on the radio flux density light curve. 
    {\it Right: } 
    Same as the left panel except that a median filtered radio flux density light curve, using a 36 s wide window, is plotted with the thermal energy evolution.  }
    \label{fig7}
\end{figure}
{The left panel of Figure \ref{fig7} shows a superposition of the \ether\ and the radio light curve for the persistent source. 
The radio light curve was obtained by averaging the SPREDS for every 4-minute observation, across the frequency axis. The light curve shows a large number of short lived impulsive features.
It is also evident that the baseline flux density of light curve is elevated during the flare period. This is likely due to the blending of numerous episodes of impulsive emissions which remain unresolved in these observations.
For a more robust study of the nature of variations in radio emission we examined the mean filtered radio light curve using a window length of 10 s (left panel).
This revealed the presence of quasi-periodic oscillations (QPOs) in the light curve.
These oscillations are of larger amplitude during the flare as compared to the pre-flare and the decline phases. 
The right panel shows the running median filtered radio light curve obtained using a 36 s wide window.
As this window is much larger than the period of the QPOs, but much smaller than the time scale of the flare itself, it is a good choice for estimating the evolution of the baseline radio flux density, and comparing it to the \ether\ light curve.
{The very similar temporal evolution of these two physically very different quantities provides a strong evidence that the sites of the type-I noise storm emission and the ARTB associated with loop heating are being powered by the same underlying source. } 
}

Note that the data used to generate the radio light curve are centred at different, but nearby frequencies.
Using the Z density model chosen for this work, the difference in the coronal heights corresponding to the two central frequencies used is $0.03$ \Rsun ($\approx 20\ Mm$), i.e. less than 10\% of the {pressure} scale height in the region.
%Assuming a typical density model of the corona, say the one by \citep{zucca2014}, we find that the height corresponding to the frequency difference is $0.03$ \Rsun, which is less than 10 \% of the scale height in the region. 
Therefore, these observation bands probe practically the same coronal region. This is also evident in the continuity of the radio flux density light curve across observation boundaries. 

\subsection{X-ray analysis}
\begin{figure}
  \centering
  \includegraphics[width=18cm,height=6cm]{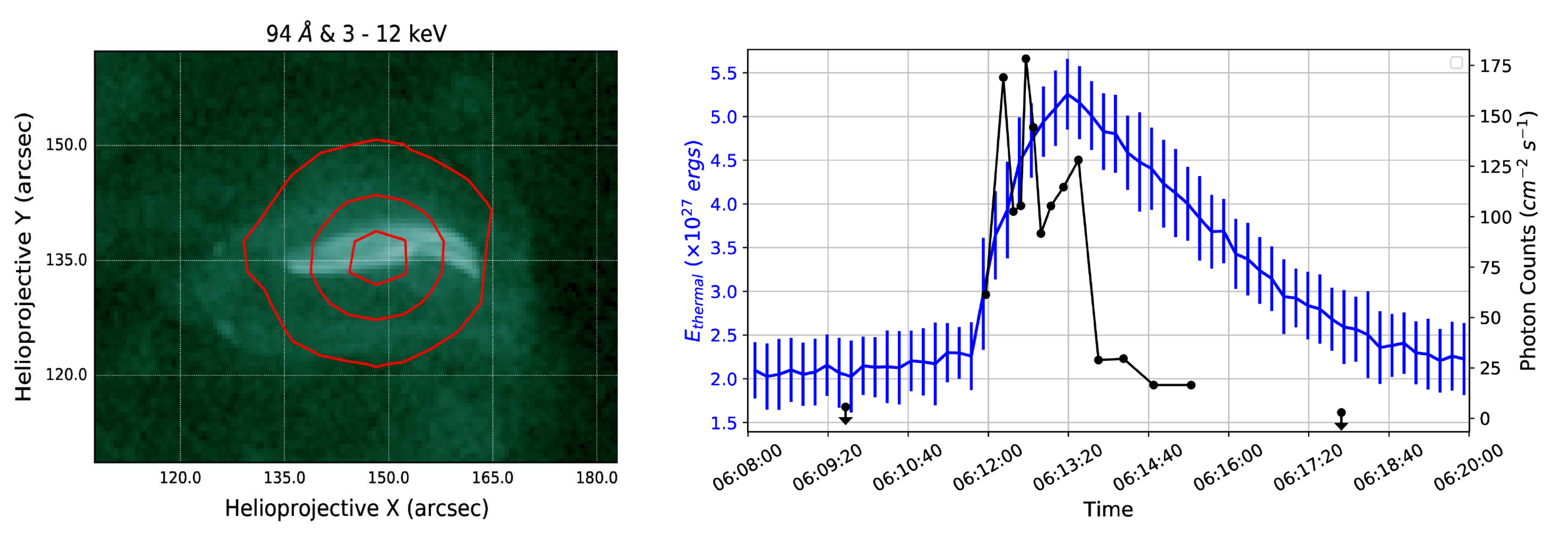}
  \caption{
  {{\it Left:} AIA 94 \AA\ image of the bright coronal loop of interest obtained by averaging five images around 06:13:00 UT.
  Superposed are the contours of the corresponding RHESSI image, obtained by averaging over 10 s around 06:13:03 UT and across the 3 -- 12 keV bands.
  Contour levels are at 30, 60 and 90 \% of the peak.
  {\it Right:} A superposition of the light curves for \ether\ in blue and for the RHESSI source flux in black.  All these RHESSI fluxes are obtained as the flux enclosed within a contour at 60\% of the peak flux. The first and the last data points in the RHESSI light curve marked with a downward arrow are upper limits to the X-ray flux in the region during the times of non-detection.
  }}
  \label{fig8}
\end{figure}

As discussed in Section \ref{obs}, the flare time is marked by a rise in the low-energy X-ray bands of RHESSI (3 -- 6 and 6 -- 12 keV; Fig.\ref{fig1}).
{The observed rise in emission has contributions from two sources of comparable flux,} one associated with the bright loop of interest and the other with a weak active region jet associated with the type-III bursts.
{In order to unambiguously obtain the light curve for the X-ray source associated with the bright loop, we imaged this source.} The imaging analysis combined the RHESSI channels from 3 -- 12 keV. 
Based on the detector performance during the observation period, the data from 3F, 6F, 8F, and 9F detectors were only used. 
Imaging was done using the CLEAN algorithm. 
While imaging, we attempted to obtain the finest time resolution possible while maintaining a reasonable signal-to-noise ratio ($SNR$). 
We could attain $SNR > 10$ with time averaging as low as 5 s during the peak time of the flare. 
Attempts to image this region before the start and after the end of the RHESSI flare, did not lead to a source detection even with several minutes of averaging. 
{The left panel of Figure \ref{fig8} shows a superposition of the X-ray source contours from the time it was at its peak brightness, on an AIA 94 \AA\ image around the same time.
This clearly establishes the co-spatial nature of the soft X-ray source with the bright EUV loop.
Immediately after the start of the rise phase of \ether, the soft X-ray source appeared on the bright loop centred at  ($150^{\prime\prime},136^{\prime\prime}$).
Right panel shows the evolution of the total flux enclosed within the 60\% contour in the RHESSI images, along with the evolution of \ether. The first and the last points in the RHESSI light curve are obtained for the cases of non-detection, where the X-ray maps remained noisy even after averaging the data over 4 minutes. Hence, these estimates provide an upper limit for the X-ray flux in the region in the pre-flare and post-flare period. 
The bright RHESSI source which appeared close to 06:12:00 UT attained its peak flux about 50 s before the observed \ether\ peak and faded away rapidly, falling below detection threshold about 2 min beyond the \ether\ peak. The co-temporal nature of the \ether\ and the X-ray flux evolution is self evident. 
}
\subsection{HMI data analysis}\label{hmi}
{We use the HMI vector magnetogram data to reconstruct the coronal magnetic field. 
The values of magnetic field components at the photosphere provided by HMI are used as the boundary conditions and the magnetic field is assumed to be force-free.}
The relevant HMI magnetogram data are given by Space weather HMI Active Region Patch (SHARP) for Active Region (AR) 12203, 2014 November 3, 06:12 UT. The SHARP patch is $608\times 309$ pixels in size, where each pixel is 0.5 arcsec across, and the highest value of the radial field component at the photosphere is $2.48\times 10^3$ G. The nonlinear force-free field (NLFFF) code CFIT was used to solve the NLFFF equations using Grad-Rubin iteration (\citealp{wheatland07_nlfff}). CFIT works in a Cartesian geometry, so we approximate the SHARP field strength data, given in a Lambert cylindrical equal-area projection, as field values on a Cartesian grid and solve for the magnetic field in a Cartesian box, where the photosphere is located on the $x$-$y$ plane at $z=0$ and the positive $z$-direction is the radial direction away from the Sun (in other words, we take $B_\phi=B_x$, $B_\theta=-B_y$, and $B_r=B_z$). 

The Grad-Rubin method solves for the magnetic field and the force-free parameter, $\alpha$, iteratively, using the value of the normal component of the magnetic field and the value of $\alpha$ at the photosphere as boundary conditions. The other boundary conditions we impose on the solution box are that the field decays to zero as $z\rightarrow\infty$ and that the field is periodic at the sides of the box. Open field lines, i.e., those that cross the boundaries of the box either at the sides or the top, are assigned $\alpha=0$. Values of $\alpha$ on the photosphere are obtained from the local vertical component of the current density, which are calculated from the derivatives of the horizontal components of the magnetic field. In general, HMI magnetogram data over-prescribe the problem by giving values of $\alpha$ over both positive and negative polarities of the active region. Therefore, two solutions can be constructed, depending on which $\alpha$ values are used as boundary conditions, those from the positive (P solution) or the negative magnetic polarity region (N solution). 

In practice, P and N solutions can be very different, because the vector magnetogram data are inconsistent with the force-free assumption (\citealp{DeRosa09_nlfff_CodeAssess}). To address this problem, \citet{Wheatland09_nlfff_implemntActReg} proposed the ``self-consistency procedure". P and N solutions are separately calculated, then the values of $\alpha$ from the two solutions are averaged, weighted by the associated uncertainties. At the end of one such self-consistency ``cycle", we arrive at a new set of boundary conditions for $\alpha$. P and N solutions are then calculated again from the new set of boundary conditions. The process is repeated until the P and N solutions agree and we arrive at a self-consistent NLFFF model.

\begin{figure}
  \centering
  \includegraphics[width=\textwidth]{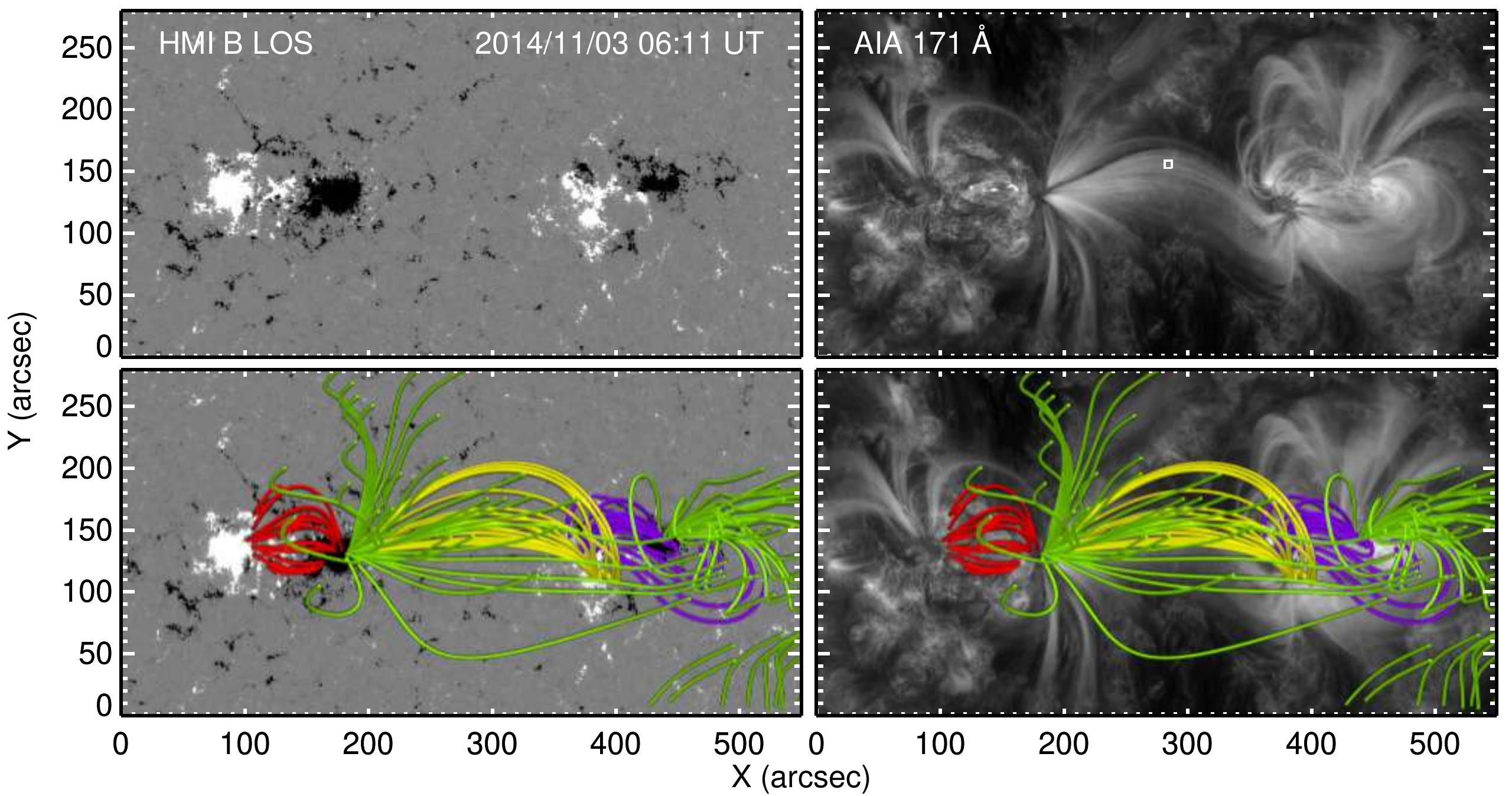}  
  \caption{{\it Top Left: }HMI magnetogram showing the sunspot group associated with the bright loop is visible to the left. 
  A large loop structure bridges this sunspot group with another sunspot group to its right. {\it Top Right: } The same region as seen in AIA 171 \AA\ map. The bright loop region centred around ($150^{\prime\prime},130^{\prime\prime}$) is relatively dim in this AIA band which is sensitive primarily to plasma at $0.65\ MK$. The small square marked at ($185^{\prime\prime},155^{\prime\prime}$) marks the apex of the large loop. {\it Bottom: } Extrapolated magnetic field structures are overlaid on the HMI ({\it Left}) and AIA 171 \AA\ map ({\it Right}). The field lines are shown in different colors to highlight the  difference in their topology and footpoint connectivity.}
  \label{fig9}
\end{figure}

Initially, we applied CFIT, with the self-consistency procedure implemented, to the the active region containing the bright loop (AR 12203), 2014 November 3, 06:12 UT SHARP magnetogram. However, we were unable to reconstruct the large loop structure, clearly visible in EUV, from this SHARP patch alone. We extended the magnetogram to include the neighbouring active region, AR 12202. The magnetogram, now $1367 \times 600$ pixels in size (each 0.5 arcsec across), was rebinned by a factor of 2 to speed up calculations, so that our solution box size is $683\times 300\times 300$ pixels, with each pixel being 1.005 arcsec. The top of the solution box is, therefore, at $1.314\ $\Rsun. In addition, to deal with noise in the data, we censor currents on the boundary surface with signal-to-noise ratios less than 1, i.e., we set $\alpha=0$ wherever SNR($J_z$) < 1. We also censor out weak field regions, i.e., we set $\alpha=0$ where $|B_z|<0.05 \mathrm{max}(|B_z|)$. 

The resulting coronal field extrapolation is presented in Figure \ref{fig9}. The large loop structure connecting AR 12203 to AR 12202 is well replicated by the model (the yellow field lines). The apex of the large loop connecting the two regions is at $\approx 1.14\ $\Rsun\ and the magnetic field strength at this height is about 5 G. The available free energy in the magnetic field close to the bright loop region was estimated from the difference between the total magnetic energy in the region and the energy of the potential field. This value turned out to be $\approx 5.1 \times 10^{29}$ erg, which is about two orders of magnitude larger than the thermal energy in the region.

\section{Discussion}\label{discussion}
%--Structures in SPREDS\\
{The previous section presented a multi-wavelength imaging analysis of a bright loop region using EUV, 
X-ray and metrewave observations. The coronal magnetic field extrapolation done using the NLFFF technique on the vector magnetogram data was also presented. 
The EUV and X-ray analysis revealed an ARTB event during our observations. The emission features in the SPREDS for the associated persistent metrewave source were typical of a type-I noise storm.
The ARTB was associated with a microflare. 
The rise in \ether\ of the bright loop region during the flare is accompanied by an enhancement in the average radio flux density, and as \ether\ falls to its pre-flare value, the radio flux density also drops (Fig. \ref{fig7}). 
These observations show that the radio and high energy events are co-temporal. There is also clear evidence for the spatial association of the sources. The noise storm source is seen slightly shifted by $\approx 1^\prime$ to the right of the bright loop region and it overlaps the large loop structure connecting AR 12203 and AR 12202. 
Though the sky location of the radio source is observed to fluctuate around its mean, the vast majority of these fluctuations are below 25\% of the FWHM of synthesised beams at the respective observation frequencies. Such fluctuations in the observed source location are expected to arise due to radiowave scattering \citep[e.g.][]{steinberg1971,kontar2017,patrick2018_densmodel_frmtypIII,Atul19_QPO}.
The observed asymmetry in the distribution of the location of the compact radio source (Fig. \ref{fig6a}) is likely due to the anisotropy in the distribution of coronal density inhomogenities encountered by the rays as they traverse out of the corona \citep{robinson_scat1983}.
The magnetic field extrapolation (Sec. \ref{hmi}) revealed that the peak of the large loop structure extends to the height from where the radio emission is expected to arise from, and share a photospheric foot point with the bright loop region. 
%Together, these constitute the evidence for a spatio-temporal association and a dynamical link between the radio with the high energy sources.
Together, these analyses establish the spatio-temporal association and the dynamical link between the emission regions seen in the multi-waveband images.}
%The associated radio emission comes from a source located much higher in the corona, and is co-spatial with the large loop structure connecting AR 12203 and AR 12202.
%The radio source is likely to be associated with the large loop structure connecting AR 12203 and AR 12202 and extending up to a heliocentric height of about 1.14 \Rsun.
%These are all manifestations of an interconnected system driven by the same underlying source. 

This section examines the co-evolution of the 
observed emissions at different wavebands to put together a physical picture for the mechanism driving them. We start by noting that flares are inherently non-equilibrium processes, while the thermodynamic quantities determined using the DEM analysis rely on thermal equilibrium physics. These quantities (Equations \ref{eqn4} -- \ref{eqn6}) therefore cannot generally be interpreted in their usual thermodynamic sense, but should rather be regarded as equivalent thermodynamic quantities, which are proxies for various properties of the non-equilibrium particle energy/velocity distribution. 
For instance, a rise in the the local average temperature and the energy represent an increase in the average velocity dispersion and the free energy respectively, due to the introduction of supra-thermal electrons in the region. 

Starting around 06:11:40 UT, the quantities determined using the $DEM$ analysis, $\langle T \rangle$, \ether\ and $\langle n_e \rangle$, show a rapid rise from their pre-flare equilibrium values of $4\ MK$, $2 \times 10^{27}\ erg$ and $8 \times 10^9\ cm^{-3}$ to attain the respective peak values of about $6\ MK$, $5.2 \times 10^{27}\ erg$, and $1.4\times10^{10}\ cm^{-3}$. 
This rapid rise phase ends at 06:13:19 UT and beyond this, these quantities slowly relax to their pre-flare values. 
The rising phase coincides exactly with the presence of a co-spatial RHESSI source (Fig. \ref{fig8}).
The appearance of this source at the start of the heating phase and its dramatic decline below the detection threshold past this phase implies that some heating mechanism was active for this duration, of which the RHESSI source was a consequence. 
The energy required for the loop heating likely came from the local magnetic field. 
As mentioned in Section \ref{hmi}, the magnetic free energy in the region, being two orders of magnitude larger than the generated heat energy, is capable of powering the heating.
%In Section \ref{hmi} the magnetic free energy in the bright loop region was estimated to be about $5.1 \times 10^{29}$ erg. This is an order of magnitude larger than the energy rise observed in the loop (Fig. \ref{fig5}) and, hence, can easily supply the energy needed for heating the loop. 
%A similar behaviour is seen for the density as well. 
\subsection{Estimating the heating rate}\label{sec:heat_rate}
The quantity $\delta E = \langle E_{peak-flare} \rangle - \langle E_{pre-flare} \rangle$ provides an estimate of the excess free energy which was deposited in the loop during this flare.
{It is estimated to be $\approx 3.2 \times 10^{27}$ erg} and is consistent with the typical energy budget of a microflare (\citealp{Asch_2000_flarestats_n_regimes,Ash}).
%This analysis estimates the excess free energy, $\delta E = \langle E_{peak-flare} \rangle - \langle E_{pre-flare} \rangle$, which was deposited in the corona during this flare to be $10^{28} erg$, which is consistent with the typical energy budget of a mircoflare (\citealp{Ash}).
It is useful to check if this excess free energy can indeed heat the loop plasma to a kinetic temperature of $6$ MK, within the observed period of about $100$ s.
%We investigated if this excess free energy can indeed heat the coronal plasma to a kinetic temperature of $6 MK$, within the observed period of about 2 minutes based on the model proposed by \citet{RTV1978} (RTV law hereafter). 
For this we use the model proposed by \citet{RTV1978} (hereafter, RTV model), which describes the thermal energy balance in the corona, assuming an equilibrium between heating and cooling rates. 
This model accounts for conductive and radiative heat losses using approximate power-law models. It neglects the spatial variation in the heating scale height and gravitational stratification of the loop, though these effects become important when the length of loop strands exceed the respective scale heights. 
Under these conditions, the RTV model provides an expression for the equilibrium heating rate, $\dot E_H$, required to maintain a steady temperature profile across a coronal loop segment of length $L$ and volume $V$, with a maximum temperature of $T$ at the loop top.
\begin{equation}
\dot E_H = 0.95 \times 10^{-6} {L}^{-2}T^{7/2}V\ {\textrm{erg s}}^{-1}, \label{rtvHeqn}
\end{equation}
The segment length $L$ is taken as the half length of the loop for symmetric loops. 
The RTV model assumes the equilibrium heating/cooling rate, to be uniform along the loop. 
The scaling law (Eqn. \ref{rtvHeqn}) derived using the model, referred to as the RTV scaling law, can be used to get a zeroth order estimate for the equilibrium heating/cooling rate. It is evident that in the pre-flare phase, the loop was being maintained at an equilibrium temperature of $4$ MK (Fig. \ref{fig5}). 
At the start of the flare, the heating rate suddenly increased to a larger value, leading to the observed rise in X-ray flux from the loop region and a rise in $\langle T \rangle$ deduced from the DEM analysis.
The peak value of $\langle T \rangle$ is determined to be $6$ MK.
The simplest model for a heating source which one can envisage is that of a source whose heating rate suddenly increased to a value which remained steady throughout the heating phase.
For such a model, the minimum value of enhanced heating rate required has to be sufficient to raise the  temperature of the loop to $6$ MK.
Using a loop length of $22$ Mm and $V \approx 1.58 \times 10^{26}$ cm$^{3}$ (Sec. \ref{DEM}), and assuming the RTV scaling law to provide approximate estimates under these conditions, we estimate this enhanced heating rate, $\dot E_H$ to be  $6.35 \times 10^{25}$ erg s$^{-1}$.  
We assume that the cooling efficiency remained at the pre-flare value, resulting in a sudden dumping of excess thermal energy in the loop. Considering the pre-flare equilibrium temperature of 4 MK as $T$, assuming that the spatial variation of temperature is negligible, we  estimate the typical cooling rate using Equation \ref{rtvHeqn}, invoking the fact that there existed an equilibrium between heating and cooling rates in the pre-flare phase. The cooling rate $\dot E_C = 1.59 \times 10^{25}$ erg s$^{-1}$. 
Then the typical heating time, $t_{heat}$, required to heat the plasma to $6$ MK by the enhanced heating source would be $\approx \delta E/(\dot E_H - \dot E_C$), which is about $67$ s. Since this estimate is a strong function of loop length, to get a rough upper limit of the heating time, the loop length of $32$ Mm, which was obtained by a assuming semi-circular loop geometry, can be used. {This results in a $t_{heat} \approx 142$ s.} 
For further analysis, a typical value of 25 Mm will be used for the loop length. The estimate of $\dot E_C$ for this loop length is about $1.2\times 10^{25}$ erg s$^{-1}$ and the $t_{heat}$ evaluates to around 86 s.
In reality, $\dot E_C$ depends on the instantaneous effective temperature. So, as the loop gets heated up, its cooling efficiency should also increase.
Hence, the time scales computed above are lower limits to $t_{heat}$ for different loop lengths.
{The observed $t_{heat}$ of $100$ s lies within the estimated range of lower limits ($\approx 67$ -- 142 s).} Note that the $\dot E_H$ computed using $T = 6\ MK$ is a lower limit for any given loop length. In reality the enhanced heating power could be much higher  resulting in a lower $t_{heat}$ estimate.
Given the uncertainty associated with the loop length estimate, $\dot E_H$, the approximate nature of the RTV scaling laws and that these estimates represent lower limits, the agreement with the observation suggests that it is plausible for such a heating source to raise the temperature of this loop to $6$ MK within the observed duration. 

Alternatively, an effective cooling rate was estimated using the observed duration of the heating phase of the flare ($t_{heat} = 100$ s), the $\delta E$ estimated using DEM analysis, and assuming the $\dot E_H$ value for $T = 6\ MK$, used above. As, $t_{heat} = \delta E/(\dot E_H - \dot E_C)$, the only unknown in this expression now is $\dot E_C$.
$\dot E_C$ evaluates to $1.7\times10^{25}$ erg s$^{-1}$, assuming a loop length of $25$ Mm.
Interestingly, this value of $\dot E_C$ is close to that obtained for this loop length in the pre-flare case ($1.2 \times 10^{25}$ erg s$^{-1}$). It also corresponds to an effective temperature of $4.4$ MK, which is not far from the equilibrium pre-flare value.
This suggests that, the assumption in the heating model that the cooling rate remained steady at the pre-flare value during the heating phase, is not unreasonable. 

\subsection{Estimating the cooling time}
As seen in the right panel of Fig. \ref{fig8}, the soft X-ray source suddenly drops below the detection threshold very close to when \ether\ peaks.
We regard this as an evidence for the switching off of the enhanced heating rate responsible for the temperature rise, and assume the heating rate to have dropped to the pre-flare level at this point in time. By this time, as we had computed in the previous subsection, the cooling rate would have risen by a small amount from the pre-flare value. 
The loop then goes through a non-equilibrium phase where the rate of heat loss via radiative and conductive pathways is larger than the heating rate and the temperature of the loop begins to drop. 
Here we estimate the cooling time, $t_{cool}$, required for the loop to release the excess heat energy $\delta E$ deposited in it and relax to its pre-flare equilibrium. 
This timescale is given by, $t_{cool} = \delta E/(\dot E_C - \dot E_H)$, where $\dot E_H$ and $\dot E_C$ now correspond to the cooling phase of the flare.
Based on the assumption of sudden switching off of heating source beyond the peak time of the flare and the loop relaxing to its pre-flare state, the $\dot E_H$ in this regime corresponds to that required to maintain the plasma at $4\ MK$.
Assuming to zeroth order that $\dot E_C$ remains a constant during the cooling phase, we use the estimate of the slightly enhanced effective cooling rate computed at the end of the previous subsection.
Using these estimates of $\delta E$, $\dot E_H$ and $\dot E_C$, {we estimated the cooling time $t_{cool} = \delta E/(\dot E_C - \dot E_H)$ to be 655 s.}  
Our observation window extends only to $400\ s$ beyond the peak in \ether\ and all three quantities, $\langle T \rangle$, \ether\ and $\langle n_e \rangle$ approach their pre-flare levels by the end of the observing duration. 
This is consistent with the expectation based on the $t_{cool}$ estimate.
This demonstrates that both the heating and cooling time scales are consistent with the abrupt switching on and off of a heating source, as suggested by the synchronised appearance of the soft X-ray source at the loop location.

\subsection{Implications  of the non-thermal radio emissions}
\label{subsec:radio-emissions}
{As mentioned at the beginning of Section \ref{analysis}, the minimum coronal height of the radio source is $1.14$ \Rsun, and the typical magnetic field at these heights was found to be around 5 G.
This leads to a gyro-frequency, $\omega_B$, of $15$ MHz which is much smaller than $\omega_p = 229$ MHz. 
As height increases in the corona, both $\omega_B$ and  $\omega_p$ fall. 
The former always remains lower than the latter, as it decreases relatively faster (\citealp{morosan16_coronal_param_trends_ECMI}).
However, for gyrosynchrotron emission to be observed and the electron cyclotron maser instability (ECMI) to operate, it is essential to have $\omega_B > \omega_p$ (\citealp{melrose82_ECME_theory, gary2004_radio_emission_regimes}).
This rules out these emission mechanisms. 
The only remaining possibility for origin of these bright fine features are plasma emission mechanisms.
These mechanisms are triggered by the two-stream instability and lead to coherent radio emission. 
This emission, produced by the passage of accelerated electron beams, is known to be generated at the local $\omega_p$ and its harmonic (\citealp{ginzburg1958,Tsytovich69,melrose1972}).
For further analysis, it is assumed that the observed emission is at the local $\omega_p$.  
Note that the following analysis also holds if the emission is assumed to be harmonic instead.

The SPREDS for the radio source shows numerous bright but short-lived and narrow-band emission features, reminiscent of type-I bursts (Fig. \ref{fig6b}). 
In the following text we refer to these features as bright fine features, which typically last for $0.5$ -- $2$ s and span about $10$ -- $13\ MHz$. These are likely to  be coherent plasma radiation triggered by electron beams propagating outwards along the magnetic fields. The radio emission is then expected to drift to lower frequencies as a consequence of the radially decreasing coronal electron density and hence $\omega_p$.
In SPREDS, however, we do not find evidence for frequency drifts in these bright fine features. 
This could simply be a consequence of the $0.5\ s$ time resolution being too coarse to be able to detect these spectral drifts.}

\begin{figure}
  \centering
  \includegraphics[width=18cm,height=6.5cm]{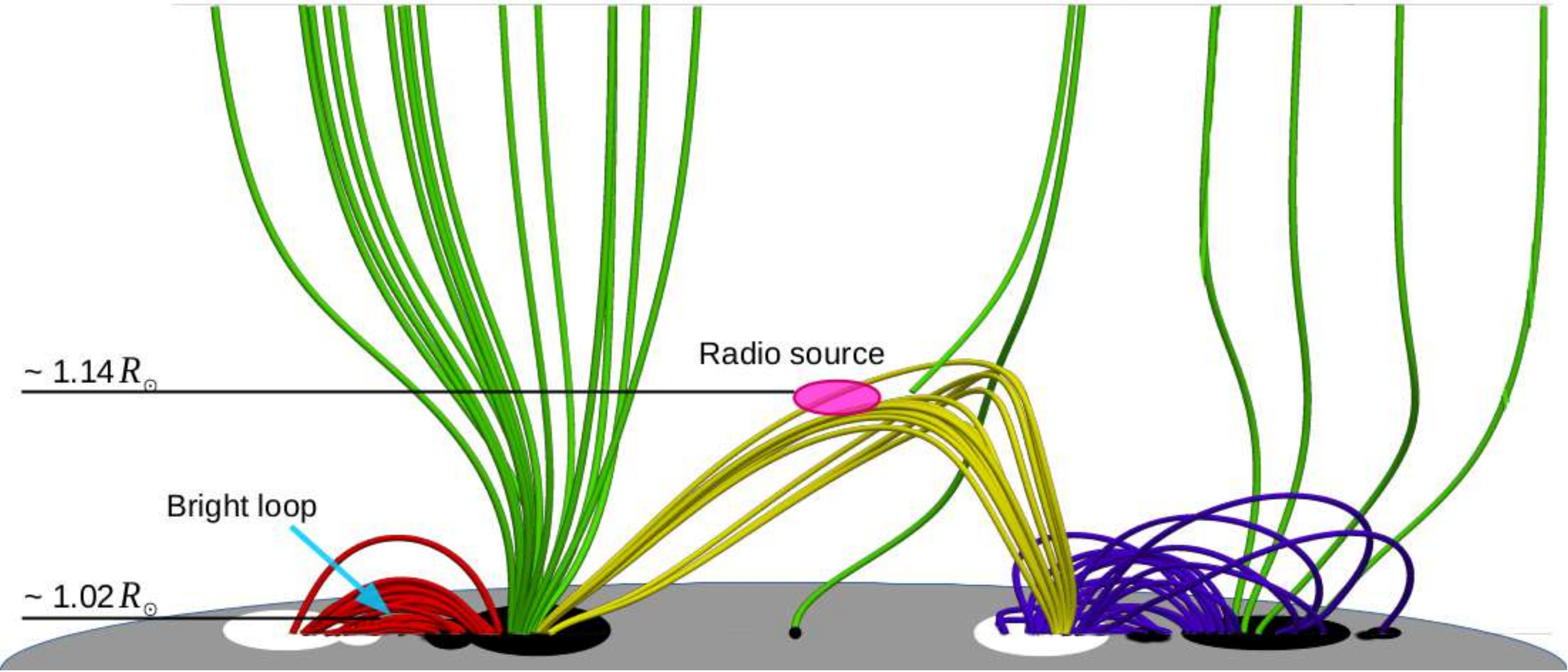}
  \caption{{NLFFF extrapolation of the coronal magnetic field. The surface on which the extrapolated field is embedded is a cartoon of the HMI magnetogram. The approximate location of the radio source at \omgp\ = 229 MHz is denoted by an ellipse. The location of the bright loop is marked by an arrow. The 15 MHz observation band corresponds to a narrow height range of $\approx 0.014$ \Rsun\ around $\approx 1.14\ R_{\odot}$.}}
  \label{fig10}
\end{figure}
\begin{figure}
  \centering
  \includegraphics[width=18.2cm,height=6cm]{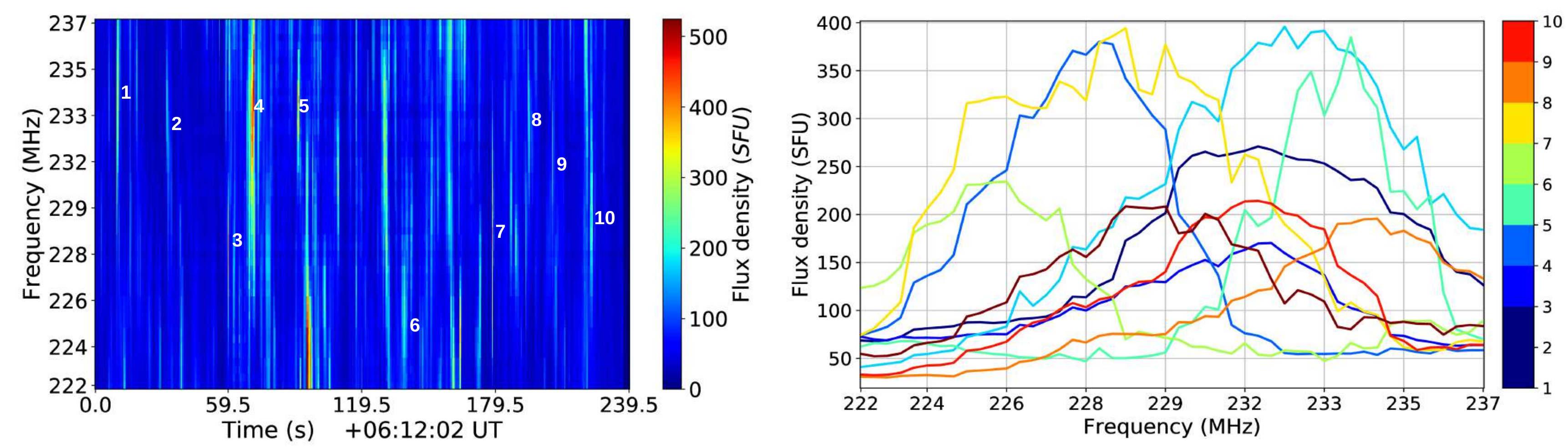}
  \caption{{\it Left: }SPREDS during the period when the X-ray source was present. Numbered are a few chosen fine features with varying intensities and frequency extent. {\it Right: } The Flux density variation of the chosen fine features across frequency at a fixed time-step. The flux density evolution shows a gradual rise and fall phase. Colorbar indicates the serial number of the feature.}
  \label{fig11}
\end{figure}

Consider the observations centred at $229\ MHz$, which overlap with the period when the X-ray source was seen in the $3$--$12\ keV$ images. The local plasma density corresponding to $\omega_p = 229\ MHz$ is $6.5\times 10^8$ cm$^{-3}$. 
{Figure \ref{fig10} illustrates the physical scenario emergent from the interpretation of the observations and the coronal magnetic field structure extrapolated using the NLFFF technique. 
The bright loop extends to a height of around 1.02 \Rsun, while the radio source is located at a height of around 1.14 \Rsun. The radio source is associated with the large loop connecting the active regions AR 12203 and AR 12202. 
The observation bandwidth of 15 MHz correspond to a very narrow height range of $\approx 10\ Mm$ ($\approx 0.014$ \Rsun), according to the Z model. 
Magnetic field strength of $5\ G$ estimated in the radio source region leads to a local \alf\ speed of $0.4\ Mm\ s^{-1}$.}
The left panel of Fig. \ref{fig11} shows the SPREDS during the period of interest. The typical spectral extent of the bright fine features ($10$ -- $13\ MHz$) corresponds to a distance range of $\approx 7$ -- $8.5$ Mm using the Z model.
The lack of a discernible spectral drift within the imaging time resolution of $0.5\ s$ provides lower limits for the speeds of electron beams, i.e $14$ -- $17$ Mm s$^{-1}$. 
The observed lifetime of these bright fine features ($\approx 0.5 $ -- $2\ s$) compares well with the ion-electron collision timescale, $\tau_{col}$. $\tau_{col}$ is estimated to be about $1.0$ -- $1.8\ s$ for a plasma with a density corresponding to $\omega_p = 229$ MHz at a temperature of $4$ -- $6$ MK.
This estimate is based on the assumption that the local temperature at the radio source is close to the effective temperature of the bright loop and was computed using the Lorentz collision model (\citealp{Krall73_plasmaBook}): 
\begin{eqnarray}
    \tau_{col} = \frac{4\pi n_e \lambda_{D}^3}{\omega_p \ln(12 \pi n_e \lambda_{D}^3)} \approx \frac{3\times 10^5 T^{3/2}}{n_e \ln(12 \pi n_e \lambda_{D}^3)},\label{tcol}
\end{eqnarray}
where $\lambda_D$ is the Debye length. 
The close match between the observed lifetime of the particle jets and the $\tau_{col}$ estimate suggests that these jets dissipate via collisions in the local plasma. 
The overall physical picture which emerges is that of the presence of numerous particle beams radiating via plasma emission mechanisms at height of $\approx 1.14$ \Rsun.  
These particle beams damp on length scales of $7$ -- $8.5\ Mm$ within time scales of $0.5$ -- $2\ s$.
A consequence is that these beams must arise locally, implying that the magnetic reconnections giving rise to them must take place at these coronal heights.
The emissions from the stronger of these particle beams are seen as individual bright fine features in the SPREDS. 

The strength of radio emission is found to be larger during the flare (Fig. \ref{fig7}). 
This increase in emission is attributed to an enhanced presence of the bright fine features.
The strength of these features is expected to follow a distribution. 
In the instances where attempts have been made to study the occurrence frequency distribution of their energies, several authors have found power law distributions with steep negative slopes, often lower than $-2$, suggesting the presence of large number of weak fine features. \citep[e.g.][]{mercier97_typI_flarestats,iwai14_typeI_finefeature_hists, mugundhan16typI_feature_hist, Suresh17_waveletbasedstudy}. 
A similar situation is likely to prevail here. This would in turn imply the presence of a large number of fine features too faint to be observed individually and too numerous to be resolved and form the baseline coherent radio flux density. During the flare, their strength and occurrence rates increased causing the observed raise in baseline flux density.

An implication of the above scenario is that particle beams arising from the reconnection sites should be bi-directional and must propagate both upwards and downwards. 
This should lead to short-lived emission features drifting to both higher and lower frequencies from their point of origin, corresponding to the beams propagating downwards and upwards respectively (\citealp{Ash93_bidirectional_typIII,Melendez99_TypeIII_stats}).
While the time resolution of the these data is insufficient to be able to see these drifts, they have sufficient frequency resolution to be able to test a different prediction of this scenario.
In addition to the drift, the intensity of the coherent emission observed should drop as these beams propagate due to collisional damping of these beams. 
Hence the spectrum of these bright fine features should have a central peak at the frequency corresponding to the $\omega_p$ of the location of the acceleration event, with intensities falling as one moves farther from this spectral peak.
The left panel of Fig. \ref{fig11} shows the SPREDS during the X-ray flare with some bright fine features numbered. 
These features were selected to span a range of flux densities and their spectra are shown in the right panel.
For the features which are largely confined to the observing bandwidth, a close examination of the SPREDS shows that their spectra indeed rise, reach a maximum and then fall as the frequency increases.
This is seen much more clearly in the spectra shown in the right panel. The small fluctuations seen in the spectra are noteworthy. Similar behaviour has been observed for type-IIIb solar bursts, and has been interpreted as arising due to the presence of density inhomogenities in the medium \citep{ellis69_typeIIIb, delanoe1975_typIIIbcharacterisation, Loi14_striae_model, mugundhan2017_dnn_est,kontar2017}. {In the radio images used for this study, the typical root mean square (rms) of the flux densities in a blank sky region far from the Sun is $\approx 0.015$ SFU. The value was arrived at by estimating this rms value for all the images used in the analysis and finding their median. 
We use this value as a representative estimate of ($1\sigma$) uncertainty in our flux densities. The fact that observed fluctuations are orders of magnitude larger than the flux density uncertainty imply that they are significant.} 

As is evident from the panels in the right column of Fig. \ref{fig6b}, the bright fine features are seen during the entire observation period, though their occurrence frequency is comparatively very high during the heating phase of the flare. 
The observations in the pre-flare and flare decline phases happen to be centred at a frequency of $199\ MHz$, but they effectively probe the same region of corona as inferred in Section \ref{radio_analysis}. 
Hence, physical parameters like $\omega_p$ and $\tau_{col}$ are not remarkably different between these observing frequencies, making the analysis and conclusions drawn earlier in this section equally applicable to these observations too.
Though the observations presented here are confined to a $15\ MHz$ band ($\approx 10\ Mm$ in height range) at any given time, the similarities between the observations centred at $199\ MHz$ and $229\ MHz$ suggest that similar situation prevails across the combined observation band.

\subsection{Quasi-periodic oscillations in radio flux density}
\begin{figure}
  \centering
  \includegraphics[width=17.4cm,height=12cm]{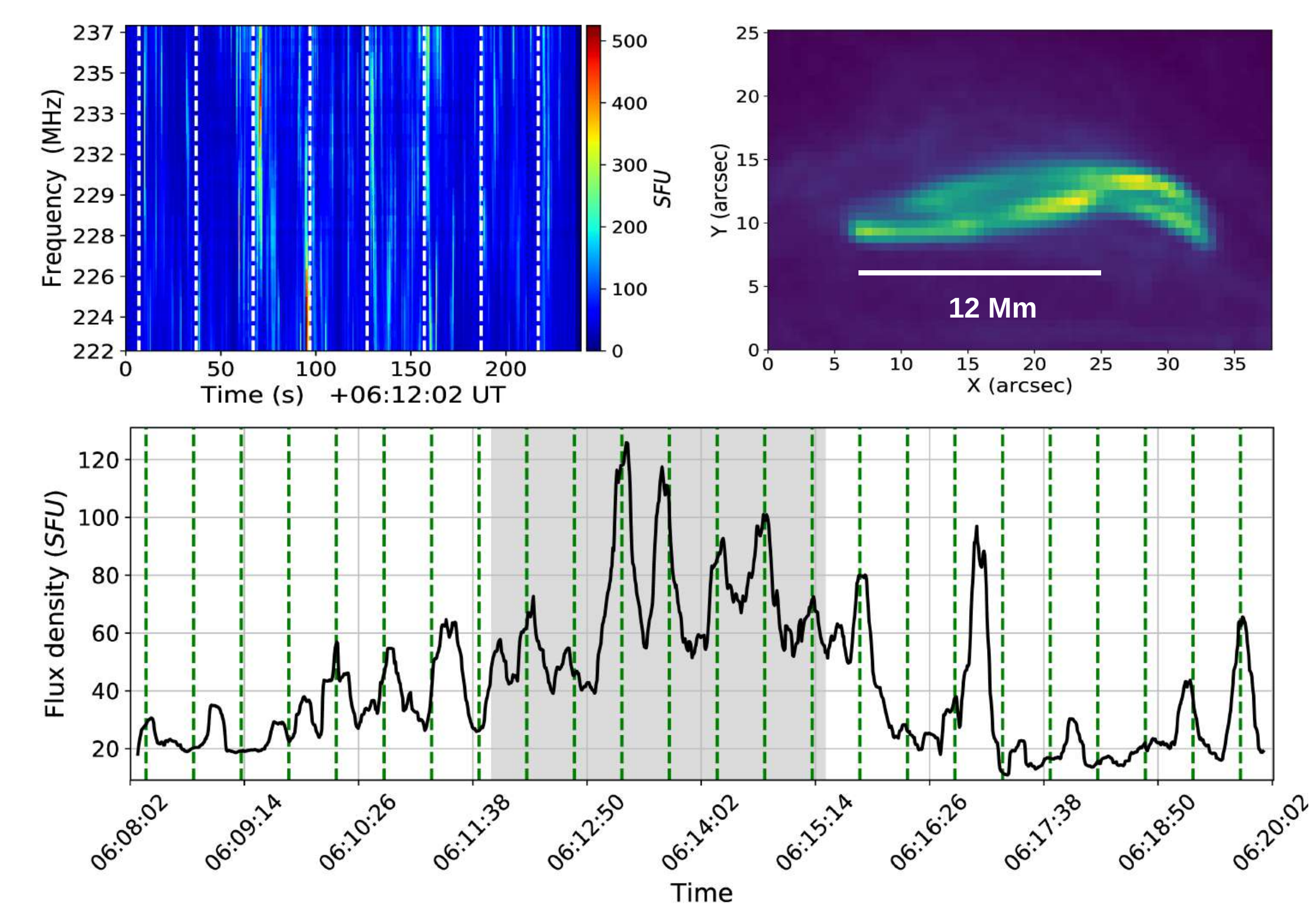}
  \caption{{\it Top: } Left panel shows the SPREDS during the  X-ray flare. Vertical dotted lines in the SPREDS plot are drawn at 30 s intervals. Right panel shows the zoomed AIA 94 \AA\ image of the bright loop region obtained by averaging four images centred around 06:13:12 UT. Two overlapping loop strands are seen with a brightening close to the cross over location. {\it Bottom: } Radio flux density light curve obtained by averaging all the three SPREDS data over the $15\ MHz$ observing band, after they have been smoothed by a running mean filter of $10\ s$ width. Shaded is the period during which the X-ray source was detected and the vertical lines are marked at 30 s intervals.}
  \label{fig12}
\end{figure}

The top left panel of Fig. \ref{fig12} shows the SPREDS corresponding to the central MWA observation, which starts close to the initiation of the heating phase and continues well into the cooling phase.
The vertical white dashed lines are drawn at  intervals of $30\ s$. 
One of these lines was drawn next to a prominent fine feature, and it is evident that practically all of them are next to one such episode of emission.
This periodicity is seen more clearly in the bottom panel of this figure. 
It shows the band-averaged radio light curve for all of the three SPREDS data smoothed by a running mean filter of 10 s temporal width, similar to the black curve in the left panel of Fig. \ref{fig7}.
The vertical green dashed lines are drawn at $30\ s$ intervals and the shaded area marks the period when the X-ray source was detectable in the $3$--$12$ keV RHESSI channels (Fig. \ref{fig8}).
These panels demonstrate there is an intrinsic preferred time scale, of about $30\ s$, at which a large number of bright fine features tend to bunch. 
The amplitudes of these oscillations are larger and they are more regular when the heating is in progress (i.e., when the soft X-ray source is seen).
Though they lose their regularity and are not as well formed, their presence is evident even during the non-heating phase.
This suggests that the mechanism driving these QPOs can operate independently of the heating process, though they become be more regular and prominent during the heating phase.
As discussed in Sec. \ref{subsec:radio-emissions}, the bright fine features are believed to be signatures of the presence of particle jets.
Hence, their quasi-periodic bunching implies a similar bunching in the excitation of these particle jets.

{It is widely believed that the magnetic foot point motions due to random convective flows at the base of the photosphere lead to a build up of magnetic stress \citep[e.g.][]{Parker72_tangential_discont, parker83_neutralSheetsEvol, parker83_braidingTheory, Parker88_nanoflares, longcope15_B_reconn_heating}.
%{\citet{longcope15_B_reconn_heating} have shown the magnetic free energy that is built up by footpoint motions dissipated via magnetic reconnection events at a rate sufficient to power local coronal heating.}
\citet{Tenerani16_fast_reconn_cur_sheets} had demonstrated that the tearing mode instabilities set up in such a system can give rise to reconnections at a rate fast enough to steadily heat the local corona.
The $30\ s$ period of the observed QPOs can be interpreted as the time scale at which the magnetic stress is built up and then relieved via fast reconnection events at the radio source height. 
Using the local \alf\ speed of $0.4\ Mm\ s^{-1}$ at the radio source region, the length scale corresponding to a duration of $30\ s$ is $12\ Mm$.
%Using the local \alf\ speed of $0.4\ Mm/s$ at the radio source region, a typical length scale can be computed across which the energy build up could have happened within $30\ s$ timescale. 
Interestingly, this length scale is very close to the twisting/braiding scale seen in the AIA 94 \AA\ image of the bright loop region (Fig. \ref{fig12}, top right panel). 
%A possible scenario which could lead to these curiously similar length scales at widely separated regions could be the following. 
Since the magnetic fields threading the radio source region and the bright loop share a common magnetic footpoint, 
they are subject to the same convective motions. Hence, it is reasonable to expect twisting or braiding in the various loops originating from the footpoint to occur at similar spatial scales ($12\ Mm$ in our case).
The time scale for a disturbance to travel across these braided structures would depend on the local \alf\ speed, which leads to a timescale of $30\ s$ at the radio source region.  
The \alf\ speed in the regions where EUV and X-ray emissions originate are close to $10\ Mm\ s^{-1}$, leading to an \alf\ travel time across a braided structure of about $1\ s$. Therefore, while periodicity is evident in the radio data, we do not expect to see periodic behavior in the EUV observations because the \alf\ travel time in the bright loop region is too short compared to the current instrumental cadence. However, our model suggests that such periodicity should exist and may be observable given higher time resolution in future instruments.

The microflare must be driven by a significant 
enhancement in the rate of magnetic stress being built up in the magnetic loops involved.
This, in turn, would lead to enhanced local heating and particle acceleration via reconnection events.
The signatures of former would be seen in X-ray (and EUV) emissions as ARTB and those for latter in radio emissions as enhanced type-I noise storm emission.
If this is indeed true, as the magnetic disturbance must travel across the loops at \alf\ speed, there must be a lag between the appearance of these signatures in X-ray and radio bands.
It is therefore interesting to look for this lag and compare it with an estimate of \alf\ travel time based on NLFFF reconstruction and the Z model.
The time difference between the epochs when X-ray source and radio source attain their respective peak flux densities is about $50\ s$.
The former peaks around 06:12:30 UT (Fig. \ref{fig8}), close to the start of the heating event, while the latter attains the peak flux density around 06:13:20 UT (Fig. \ref{fig7}). 
The \alf\ travel time, $t_{A}$, from the common footpoint region to the height of the radio source, $h$, was computed as $\int_{0}^{h} dz/v_{A}(z)$, where $v_{A}$ is the spatially varying \alf\ speed.
$t_{A}$ was estimated to be $60\ s$ and is in remarkable agreement with the observed value.

%This supports the hypothesis that the convective motions at the footpoint region could have undergone significant dynamical changes during the flaring period, sending out disturbances across the loops associated with it. 
%This disturbance could have caused the enhancement in the rate of reconnection episodes leading to rise in coherent radio flux density in the radio source region and the simultaneous heating of the bright active region loop. 
%This excess energy carried by the disturbance lead to excitation of periodic particle acceleration episodes in the large loop at local \alf\ timescale across the fundamental braiding scales or scales of tangential discontinuity. 
%This probably led to the observed $30\ s$ periodicity in the radio flux density. 
The bright loop region is associated with active region AR 12203.
It has been shown to be at a temperature higher than the ambient coronal temperature (Fig. \ref{fig5}), and is also associated with a coherent nonthermal radio source (Fig. \ref{fig6b}). 
These imply that there is some mechanism in operation which is steadily maintaining it at a higher temperature and also powering the coherent radio emission. 
The presence of $30\ s$ QPOs even before the X-ray flare implies that this too is a part of the ambient behaviour, which does not owe its existence to the flare.
Nonetheless, whatever disturbances gave rise to the flare, also lead to these QPOs becoming more energetic and regular.
In addition, the baseline coherent flux density, interpreted as arising from the superposition of numerous weak coherent emissions each arising from a weak electron jet, also show a simultaneous systematic rise. 

However, even if each of the relaxing braid was
to conform to a periodicity of 30 s, a large number of such braids relaxing independently of one another would wipe out the signature of any periodicity in our data product. Hence, the clear presence of QPOs requires the existence of a network of braided magnetic strands which relax in unison. 
A promising possibility is a cascade like process, where the relaxation of one particular braid in a magnetic loop sets off an avalanche of relaxation of braids in the interconnected strands, which are already in a near critical stressed state 
%neighbouring strands, which are already in a near critical stressed state 
\citep{vanBalle86_currentcascade,wilmot11_braidedloopheat}. Once this system relaxes, it takes about an \alf\ time scale to build up near
critical levels of magnetic stress and the process continues. This interpretation suggests that the observed emission is dominated
by the contribution from a prominent braided network which relaxes via a cascade like process giving rise to the quasi-periodic bundling of bright fine features in SPREDS. This bundling of coherent fine features at 30 s timescale manifested as the QPOs in the  band-averaged radio light curve. Contributions from a large number of magnetic strands relaxing asynchronously give rise to the enhanced
radio baseline flux density. {This physical picture lends support in favor of the idea of small scale magnetic rearrangements in the vicinity of sunspots powering radio noise storms, suggested by \cite{bentley2000_MMFs_typeI_link}.}
}
\section{Conclusion}\label{conclusion}

In this work we present a multi-waveband imaging study of an {active region transient brightening (ARTB) event} using simultaneous imaging observations in metrewave, EUV and X-ray bands. 
A bright coronal loop was seen at EUV coronal bands of the AIA.
The evolution of local temperature, density and thermal energy, obtained from DEM analysis show a fast rising phase in all of these quantities leading to a peak, which slowly relaxes back close to pre-flare levels by the end of observations.
For the duration of the heating phase, a transient X-ray source appeared in 3--12 keV low energy RHESSI bands.
The metrewave snapshot spectroscopic images showed this region to be associated with a persistent bright radio source with a mean \Tb\ $\approx 10^8$ K located at a height of around 1.14 \Rsun. {SPREDS for the source revealed its resemblance to a type-I noise storm.}
The radio flux density light curve for this source varied in consonance with the local thermodynamic properties estimated via DEM analysis.
The spatial and temporal correlations between these disparate emissions establish that the high energy and radio sources are a part of a single system powered by a common underlying mechanism. 
Magnetic field NLFFF extrapolations based on HMI vector magnetograms show large loop structures that share footpoints with the bright EUV loop region and extend to the heights from where radio emission is expected to originate, providing an independent evidence for the X-ray, EUV and metrewave source being parts of a single system. 

The energy deposited in the corona during the event was consistent with the energy budget for a typical microflare ($3.2 \times 10^{27}\ erg$). 
The energy requirement for heating the plasma in the magnetic loop was estimated to be only about 5\% of magnetic free energy content in the region.
Our rough estimates, using the RTV scaling law, suggest that it is indeed feasible for the plasma in the EUV loop to be heated from a pre-flare equilibrium value of $4\ MK$ to a peak value of $6\ MK$ over the observed duration.
Similarly, the observed cooling time is also consistent with rough expectations based on the RTV scaling law.
The SPREDS for the associated radio source shows numerous short-lived ($0.5$--$2\ s$) narrow-band ($10$--$13\ MHz$) nonthermal slivers of emission (`bright fine features').
Our analysis suggests plasma emission as the only plausible emission mechanism for these bright fine features and rules out gyrosynchrotron and ECMI mechanisms.
Each of these features, hence, must arise from an electron beam born at these coronal heights.
The durations of bright fine features are comparable to ion-electron collision time scale at these heights.
Their observed bandwidths lead to a coronal height range of $7$--$8.5\ Mm$, using a reasonable coronal electron density model.
No spectral drifts were discernible in these features, perhaps due to the $0.5\ s$ time resolution of these data. 
However, the observed spectral profiles were consistent with expectations for beams propagating both towards and away from the Sun.
The band-averaged radio light curve showed that the occurrence rate of these impulsive features increased and the observed baseline radio flux density rose while the flare was in progress.
We interpret the latter to be due to the blending in of the numerous instances of weak emissions which are present simultaneously and are too faint to be resolved individually.

Very interestingly, the radio flux density light curve shows clear QPOs with a $30\ s$ period.
They are very well formed and regular during the phase when heating is underway.
Though a little less regular, they are clearly seen in the pre-flare phase, implying that they must arise from a mechanism independent of the flare.
{Curiously, the \alf\ length scale corresponding to the QPO period, at the height of the radio source is comparable to the braiding or twisting scale seen in the EUV images of the bright loop despite their separation of $\approx 0.01$ \Rsun\ in coronal height.
We suggest that braiding/twisting occurs at the same spatial scale across this vast distance because the magnetic fields in these regions arise from a common footpoint, and are hence subject to similar convective  motions. 
We find evidence for the disturbance responsible for the flare to be travelling at the spatially varying \alf\ speed to the height of the radio emitting region.
Overall, this suggests a physical picture where a large number of coronal reconnection events happen over a variety of spatial and energy scales, asynchronously with respect to one another forming the floor of the radio light curve. 
Superposed on this, are the emissions from a dominant braided magnetic loop which quasi-periodically relaxes at the local \alf\ time scale via a cascade of reconnection events, giving rise to the periodic bunching of bright fine features in SPREDS and the observed QPOs in the band-averaged radio flux density light curve.
}

\acknowledgments
This scientific work makes use of the Murchison Radio-astronomy Observatory (MRO), operated by 
the Commonwealth Scientific and Industrial Research Organisation (CSIRO).
We acknowledge the Wajarri Yamatji people as the traditional owners of the Observatory site. 
Support for the operation of the MWA is provided by the Australian Government's National Collaborative Research Infrastructure Strategy (NCRIS), under a contract to Curtin University administered by Astronomy Australia Limited. We acknowledge the Pawsey Supercomputing Centre, which is supported by the Western Australian and Australian Governments. The SDO is a National Aeronautics and Space Administration (NASA) spacecraft, and we acknowledge the AIA and HMI science team for providing open access to data and software. This research has also made use of NASA's Astrophysics Data System (ADS) and the Virtual Solar Observatory (VSO; \citealp{Hill09}). We thank Kathy Reeves for helpful discussions related to the DEM analysis and Aimee Norton for help related to vector magnetograms from HMI. {We thank the anonymous referee for the useful suggestions that have helped improve the presentation of the paper.} AMa is supported by the Australian Research Council Discovery Project (DP160102932).

\facility{Murchison Widefield Array, SDO (AIA \& HMI), GOES, RHESSI}

\software{CASA,
SolarSoft Ware,
Python 2.7\footnote{\MYhref{https://docs.python.org/2/index.html}{https://docs.python.org/2/index.html}},
NumPy\footnote{ \MYhref{https://docs.scipy.org/doc/}{https: //docs.scipy.org/doc/}},
Astropy\footnote{\MYhref{http://docs.astropy.org/en/stable/}{http://docs.astropy.org/en/stable/}},
Matplotlib\footnote{\MYhref{http://matplotlib.org/}{http://matplotlib.org/}}
}

\bibliographystyle{yahapj}
\bibliography{allref}

%\appendix
%\section{appendix section}

\end{document}